\PassOptionsToPackage{dvipsnames,table}{xcolor}
\documentclass[conference]{IEEEtran}
\IEEEoverridecommandlockouts

\usepackage{cite}
\usepackage{amsmath,amssymb,amsfonts}
\usepackage{algorithmic}
\usepackage{graphicx}
\usepackage{textcomp}
\usepackage{xcolor}

\usepackage[hidelinks]{hyperref}

\usepackage{makecell}
\usepackage{multirow}
\usepackage{longtable}
\usepackage{float}
\usepackage{tikz}
\usepackage{booktabs}
\usepackage{longtable}

\usepackage{soul}

\usepackage{bm}

\usepackage{url}

\usepackage{amsthm}

\usepackage{comment}

\usepackage{xparse}

\ifCLASSOPTIONcompsoc
\usepackage[caption=false,font=normalsize,labelfont=sf,textfont=sf]{subfig}
\else
\usepackage[caption=false,font=footnotesize]{subfig}
\fi

\usepackage{cleveref}

\usepackage{tcolorbox}

\newcommand{\pyrolyse}{\textsc{pyrolyse}}%
\newcommand{\Pyrolyse}{\textsc{Pyrolyse}}%

\newcolumntype{L}[1]{>{\raggedright\let\newline\\\arraybackslash\hspace{0pt}}m{#1}}
\newcolumntype{C}[1]{>{\centering\let\newline\\\arraybackslash\hspace{0pt}}m{#1}}
\newcolumntype{R}[1]{>{\raggedleft\let\newline\\\arraybackslash\hspace{0pt}}m{#1}}

\usepackage{pifont}
\newcommand{\greencmark}{\textcolor{Green}{\ding{51}}}%
\newcommand{\redxmark}{\textcolor{Red}{\ding{55}}}%
\usepackage{xparse}
\ExplSyntaxOn
\NewDocumentCommand{\getenv}{om}
 {
  \sys_get_shell:nnN { kpsewhich ~ --var-value ~ #2 } { } \l_tmpa_tl
  \tl_trim_spaces:N \l_tmpa_tl
  \IfNoValueTF { #1 }
   {
    \tl_use:N \l_tmpa_tl
   }
   {
    \tl_set_eq:NN #1 \l_tmpa_tl
   }
 }
\ExplSyntaxOff

\definecolor{Gray}{gray}{0.9}
\definecolor{Dgray}{gray}{0.7}
\newcolumntype{g}{>{\columncolor{Gray}}c}
\newcolumntype{d}{>{\columncolor{Dgray}}c}
\newtcolorbox{takeaway}[2][]{%
fonttitle    = \bfseries,
title        = #1#2,
boxsep       = 0.5mm,
top          = 0.1mm,
bottom       = 0mm,
toptitle     = 0mm,
bottomtitle  = 0mm,
left         = 1mm,
right        = 1mm,
left skip    = 0mm,
right skip   = 0mm,
coltitle=darkgray,
colbacktitle=black!10!white,
colback=white
}

\newcommand{\mynote}[3]{
\fbox{\bfseries\sffamily\scriptsize#1}
{\small\textsf{\emph{\color{#3}{#2}}}}}

\newcommand{\johan}[1]{\mynote{Johan}{#1}{teal}}
\newcommand{\lucas}[1]{\mynote{Lucas}{#1}{red}}
\newcommand{\gilles}[1]{\mynote{Gilles}{#1}{blue}}
\newcommand{\matthieu}[1]{\mynote{Matthieu}{#1}{orange}}

\renewcommand{\johan}[1]{}
\renewcommand{\lucas}[1]{}
\renewcommand{\gilles}[1]{}
\renewcommand{\matthieu}[1]{}

\def\BibTeX{{\rm B\kern-.05em{\sc i\kern-.025em b}\kern-.08em
    T\kern-.1667em\lower.7ex\hbox{E}\kern-.125emX}}
\begin{document}

\title{Overlapping IPv4, IPv6, and TCP data: exploring errors, test case context, and multiple overlaps inside network stacks and NIDSes with \Pyrolyse{}}

\author{\IEEEauthorblockN{Lucas Aubard}
\IEEEauthorblockA{\textit{Inria} \\
Rennes, France \\
lucas.aubard@inria.fr}
\and
\IEEEauthorblockN{Johan Mazel}
\IEEEauthorblockA{\textit{ANSSI} \\
Paris, France \\
johan.mazel@ssi.gouv.fr}
\and
\IEEEauthorblockN{Gilles Guette}
\IEEEauthorblockA{\textit{IMT Atlantique} \\
Cesson Sévigné, France \\
gilles.guette@imt-atlantique.fr}
\and
\IEEEauthorblockN{Pierre Chifflier}
\IEEEauthorblockA{\textit{ANSSI} \\
Paris, France \\
pierre.chifflier@ssi.gouv.fr}
}

\maketitle

\begin{abstract}


IP fragmentation and TCP segmentation allow for splitting large data packets into smaller ones, e.g., for transmission across network links of limited capacity. 
These mechanisms permit complete or partial overlaps with different data on the overlapping portions. IPv4, IPv6, and TCP reassembly policies, i.e., the data chunk preferences that depend on the overlap types, differ across protocol implementations. 
This leads to vulnerabilities, as NIDSes may interpret the packet differently from the monitored host OSes.
Some NIDSes, such as Suricata or Snort, can be configured so that their policies are consistent with the monitored OSes.

The first contribution of the paper is \pyrolyse{}, an audit tool that exhaustively tests and describes the reassembly policies of various IP and TCP implementation types. 
This tool ensures that implementations reassemble overlapping chunk sequences without errors. 
The second contribution is the analysis of \pyrolyse{} artifacts.  
We first show that the reassembly policies are much more diverse than previously thought. 
Indeed, by testing all the overlap possibilities for \bm{$n\leq3$} test case chunks and different testing scenarios, we observe 15 different behaviors out of 23 tested implementations depending on the protocol.
Second, we report eight errors impacting one OS, two NIDSes, and two embedded stacks, which can lead to security issues such as NIDS pattern-matching bypass or DoS attacks. 
A CVE~\cite{cve_ours} was assigned to a NIDS error.
Finally, we show that implemented IP and TCP policies obtained through chunk pair testing are usually inconsistent with the observed triplet reassemblies.
Therefore, contrary to what they currently do, NIDSes or other network traffic analysis tools should not apply $n = 2$ pair policies when the number of overlapping chunks exceeds two.

\end{abstract}

\begin{IEEEkeywords}
IP, TCP, intrusion detection, evasion
\end{IEEEkeywords}

\section{Introduction}

A host often needs to send more data than the medium or an underlying protocol can send at once. 
IP fragmentation and TCP segmentation address this generic networking problem by chunking the original (upper-layer) data into several packets.
When chunking is used, the receiver must reassemble all the chunks to reconstruct the original data packet. 
However, this chunking mechanism can result in overlaps. 
The most frequent scenario is when a chunk is retransmitted with identical data, starting and ending at the same byte offsets. 
Nevertheless, partial overlaps can also occur, and the data in the overlapping sections may differ. 
The specifications for IPv4 and TCP (as outlined in their respective RFCs~\cite{rfc791,rfc9293}) do not prohibit data overlaps or dictate how implementations should handle them (for example, whether to prioritize data from the older chunk).
IPv6 RFC specification initially did not forbid overlaps in the first drafts, but has banned them since 2017~\cite{deering2017rfc}.

Network Intrusion Detection Systems (NIDSes) match suspicious patterns or signatures of known attacks on the reassembled flow data. 
In 1998, Ptacek and Newsham~\cite{ptacek1998insertion} introduced insertion and evasion attacks, whose goal is to desynchronize the NIDS reassembly state from the monitored hosts. 
Such desynchronizations thus allows an attacker to fool the NIDS pattern-matching functionality.
Overlapping chunks with incoherent data portions is one of the insertion/evasion strategies the authors described in the paper. 
Indeed, they noticed variations in reassembly strategies for IPv4 and TCP data across NIDSes and Operating Systems (OSes).  
In response, NIDSes proposed two countermeasures: raise alerts whenever there is an overlap or associate host IP addresses with the reassembly policies they implement.     

Later studies confirmed and enriched Ptacek and Newsham's findings. 
Notably, Novak and Sturges~\cite{novak2005target,novak2007target} found that OSes reassembled in seven (resp. six) different ways the overlapping IPv4 fragments (resp. TCP segments). 
Suricata~\cite{suricata} and Snort~\cite{roesch1999snort} NIDSes provide reassembly policy configurability~\cite{rpsuricata,rpipsnort,rptcpsnort} which is based on these works.
However, IPv4 and TCP reassembly policies of OSes have evolved, and thus, Snort and Suricata can be subject to insertion and evasion attacks when supervising hosts with recent OS versions~\cite{ourdimva25paper}. 
In addition to NIDS midpoint stacks, some works~\cite{khattak2013towards,wang2017your,bock2019geneva,wang2020symtcp} tested the circumvention of several censorship systems (CSes) with IPv4 or TCP overlapping chunks with some success. 
Apart from OSes and midpoint stacks, such as NIDSes and CSes, no other IP and TCP implementation reassembly policies were tested.  

Novak and Sturges~\cite{novak2007target} and we~\cite{ourdimva25paper} previously observed different reassemblies for the same overlap type depending on whether the overlaps were tested altogether (within a unique chunk sequence) or separately. 
The IP and TCP testing contexts are thus important; however, the authors did not explore every context aspect (e.g., different \emph{More Fragments} bit unsetting strategies when multiple rightmost fragments overlap). 
Furthermore, they described implementation reassembly policies by exhaustively testing pairs ($n=2$) of overlapping chunks.
What happens if more than $n=2$ chunks overlap?    

In this paper, we tackle the following questions: \emph{How diverse and correct are IPv4, IPv6, and TCP implementations when exploring test case context and up to $n = 3$ overlaps?} 
Corollary, \emph{can the NIDSes deduce the monitored hosts' reassembly of any n overlapping chunk sequence from their $n = 2$-based policies?}
The paper's contributions are introduced as follows:
\begin{itemize}
    \item In~\Cref{sec:pyrolyse_design}, we propose a new testing tool named \pyrolyse{}, short for Protocol bYte stReam OverLapping ambiguitY reaSsembly tEsting. 
    This generic and easily extensible tool enables one to test a diverse range of IPv4, IPv6, and TCP implementation reassembly policies regarding to overlapping chunk sequences. 
    In its current implementation, \pyrolyse{} ensures exhaustiveness for test cases of up to $n = 3$ overlapping chunks and implements 42 IP and 11 TCP testing scenarii. 
    It uses a reassembly output model for $n\leq3$ chunks to describe real-world implementations\footnote{In this study, we focus on $n \leq 3$ because we primarily want to verify if $n = 2$-based implementation reassembly policies are relevant for any $n$ to improve NIDSes.  
    Since we find inconsistencies between $n = 2$ and $n = 3$  policies (\Cref{sec:uc_reassembly_policy_generalization_temptative}) for most implementations, we don't need to test $n > 3$. 
    However, testing $n > 3$ may uncover other reassembly errors than those described in~\Cref{sec:uc_reassembly_bugs}.}. 
    \item In~\Cref{sec:uc_pair_triplet_reassembly_policies}, we describe the $n=3$-based reassembly policies of some OSes, NIDSes, embedded/IoT, unikernel, NIC, and DPDK-compatible stacks. 
    We uncover a wide diversity of reassembly policies, as almost every stack has its own policy.
    \item In~\Cref{sec:uc_reassembly_bugs}, we report the reassembly errors we encountered while testing the protocol stacks. 
    We found that one OS (OpenBSD), two NIDSes (Suricata and Snort), and two custom stacks (lwIP and mirage-tcpip) reassemble some overlap cases with errors. 
    A CVE (now patched) was attributed to the Suricata IP-related error, and OpenBSD fixed the two reported bugs. 
    \item 
    In~\Cref{sec:uc_reassembly_policy_generalization_temptative}, we show that $n = 3$ test case reassemblies cannot easily be deduced from $n = 2$-based policies. 
    To explore if the NIDSes have any chance to reassemble consistently when $n > 2$, we implement some custom algorithms that try to reassemble $n=3$ test cases using $n=2$ ones. 
    We establish that protocol implementation behavior is only deductible for 25 stacks out of 61.
\end{itemize}

\section{Background}

\begin{figure}[t!]
    \centering
    \includegraphics[width=0.35\textwidth]{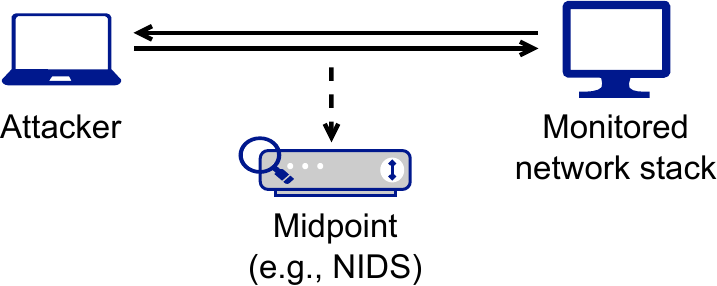}
    \caption{\label{fig:threat_model}The considered threat model to perform insertion and evasion attacks.
    }
\end{figure}

\subsection{Insertion and evasion attacks}
\label{sec:background_insertion_evasion_attacks}

In 1998, Ptacek and Newsham~\cite{ptacek1998insertion} introduced a set of IP and TCP ambiguities (i.e., a network packet that is differently process by two stacks) that may lead to the NIDS stack desynchronization with the monitored hosts.
Since the traffic monitoring machine and the monitored hosts are distinct, the NIDS receives a copy of the monitored host traffic and, thus, has no easy way to know the current host reassembly state.
\Cref{fig:threat_model} illustrates this generic insertion and evasion attack-related threat model.  
By carefully crafting some packets, an attacker can use a reassembly ambiguity to attack the NIDS or the host it monitores by \emph{inserting} or \emph{evading} malicious data in one of the two data flows~\cite{ptacek1998insertion,ourdimva25paper,wang2020symtcp}.
The attack is an insertion if the target is the NIDS and an evasion if the target is the monitored host~\cite{ourdimva25paper}. 
Some vulnerabilities~\cite{suricatacve2,suricatacve1,amosyscve} related to the NIDS stack desynchronization were recently attributed high or critical scores, meaning that these kinds of attacks are still a security concern for NIDSes. 
CSes operate roughly in the same way as NIDSes, and some works described new insertion and evasion techniques throughout the years~\cite{khattak2013towards,li2017lib,wang2017your,wang2020symtcp,wang2021themis,bock2019geneva,bock2020come,bock2021even}.

\subsection{Data overlapping ambiguity}
\label{sec:background_data_overlapping_ambiguity}

One of the IP and TCP ambiguities outlined by Ptacek and Newsham~\cite{ptacek1998insertion} consists of chunking an original data packet into several pieces and making some data portions overlap. 
Indeed, they observed that the OSes and NIDSes they tested reassembled differently the few overlapping chunk sequences under analysis, which was still recently observed~\cite{ourdimva25paper}.
An attacker can use the NIDS misassembly to insert or evade some malicious data. 
The rightmost column of~\Cref{tables/allen_relations} illustrates the overlap ambiguity. 
Evasion occurs if the monitored and monitoring flow data are respectively reassembled as ``ATTACK"/``AT00CK" or ``ATTACK"/$\varnothing$\footnote{$\varnothing$ means that the overlapping chunk sequence is ignored.}. 
Conversely, insertion occurs if the NIDS reassemble with the malicious data.

\subsection{Stack reassembly policies}
\label{sec:background_stack_reassembly_policies}

\subsubsection{$n = 2$-based reassembly policies}
\label{sec:background_stack_reassembly_policies_n_2}

\renewcommand{\arraystretch}{0.8}
\begin{table}[t!]
    \setlength\tabcolsep{3pt}
    \centering
    
    \caption{\label{tables/allen_relations}Allen's interval algebra relations.
    }
        
    \begin{tabular}{l | l  c | c  r | c}
        \toprule
        \textbf{Descr.} & \multicolumn{2}{c|}{\textbf{\makecell{Relation \textcolor{Red}{$\mathcal{R}$}}}}    &    \multicolumn{2}{c|}{\textbf{\makecell{Relation \textcolor{Red}{$\mathcal{R}$ i}nverse}}} & \textbf{Example for \textcolor{Red}{$\mathcal{R}$}}\\
        \midrule
        Meet & X \textcolor{Red}{$M$} Y  & \raisebox{-6pt}{\includegraphics[width=0.06\textwidth]{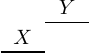}}    & \raisebox{-6pt}{\includegraphics[width=0.06\textwidth]{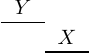}} & X \textcolor{Red}{$Mi$} Y 
        \\

        \hline

        Before & X \textcolor{Red}{$B$} Y  & \raisebox{-8pt}{\includegraphics[width=0.06\textwidth]{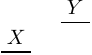}}  &  \raisebox{-8pt}{\includegraphics[width=0.06\textwidth]{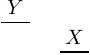}} & X \textcolor{Red}{$Bi$} Y \\
        \hline
        \hline
        Equal & X \textcolor{Red}{$Eq$} Y  & \raisebox{-6pt}{\includegraphics[width=0.06\textwidth]{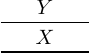}}  &   & - &
        \raisebox{-12pt}{\includegraphics[width=0.1\textwidth]{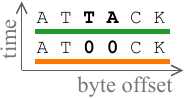}} 
        \\
        \hline

        Overlap & X \textcolor{Red}{$O$} Y  & \raisebox{-6pt}{\includegraphics[width=0.06\textwidth]{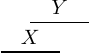}}  &  \raisebox{-6pt}{\includegraphics[width=0.06\textwidth]{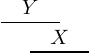}} & X \textcolor{Red}{$Oi$} Y &
        \raisebox{-12pt}{\includegraphics[width=0.1\textwidth]{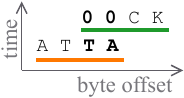}} 
        \\

        \hline

        Start & X \textcolor{Red}{$S$} Y  & \raisebox{-6pt}{\includegraphics[width=0.06\textwidth]{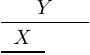}}  &  \raisebox{-6pt}{\includegraphics[width=0.06\textwidth]{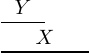}} & X \textcolor{Red}{$Si$} Y &
        \raisebox{-12pt}{\includegraphics[width=0.1\textwidth]{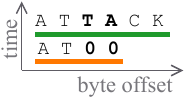}} 
        \\
        \hline

        During & X \textcolor{Red}{$D$} Y  & \raisebox{-6pt}{\includegraphics[width=0.06\textwidth]{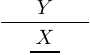}}  & \raisebox{-6pt}{\includegraphics[width=0.06\textwidth]{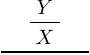}} & X \textcolor{Red}{$Di$} Y &
        \raisebox{-12pt}{\includegraphics[width=0.1\textwidth]{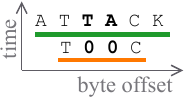}} 
        \\
        \hline

        Finish & X \textcolor{Red}{$F$} Y  & \raisebox{-6pt}{\includegraphics[width=0.06\textwidth]{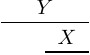}}  & \raisebox{-6pt}{\includegraphics[width=0.06\textwidth]{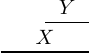}} & X \textcolor{Red}{$Fi$} Y &
        \raisebox{-12pt}{\includegraphics[width=0.1\textwidth]{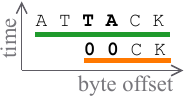}} 
        \\

        \bottomrule
    \end{tabular}

\end{table}

Suricata and Snort NIDSes implemented various reassembly policies to address the overlap reassembly ambiguity with the host they monitor. 
The users have the option to associate monitored host IP addresses with a reassembly policy. 
The implemented policies are based on Novak and Sturges' works~\cite{novak2005target,novak2007target} published in 2005 and 2007.  
The OS reassembly policies, which were obtained by exhaustively testing all the possible overlap types for $n=2$ chunks, 
have since evolved~\cite{ourdimva25paper}.
Both works observed different reassemblies when overlaps were tested individually (within multiple chunk sequences) or altogether (within one chunk sequence). 
This aspect is named \emph{testing mode} in~\cite{ourdimva25paper}.

This last work also introduced Allen's interval algebra, a spatio-temporal reasoning, to model overlapping chunk sequences for $n=2$.
The algebra comprises nine overlapping and four non-overlapping relations, as described in~\Cref{tables/allen_relations}.  
They described the reassembly policies timewisely.
An implementation may favor the oldest or newest data chunk or completely ignore the chunks for a given overlap relation.

\subsubsection{Beyond $n = 2$-based reassembly policies}
\label{sec:background_stack_reassembly_policies_n_2_beyond}

Atlasis~\cite{atlasis2012attacking} described IPv6 overlap reassembly policies and used combinations of three fragments for the testing.
The author introduced a test case generation strategy that yielded one or two distinct overlapping relations inside every fragment sequence.
Triple overlaps, i.e., a similar data portion overlaps three chunks as in~\Cref{figures/triplet}, are not addressed. 
Therefore, this strategy is not exhaustive for $n=3$ chunks. 

In contrast, Di Paolo et al.~\cite{di2023new} used a fuzzing-like approach to check OS conformity with the IPv6 specification~\cite{deering2017rfc} when processing overlapping data fragments.
They derived the Shankar and Paxson model~\cite{shankar2003active} by permuting and duplicating the chunks and found that none of the tested OSes (Linux, Windows, or BSD-based) comply. 
In their second and most exhaustive testing scenario, they performed the permutation and four duplications of the original fragment sequence, which is composed of six fragments. 
In total, 30 fragments overlap with $O$, $Oi$, and $Eq$ relations. 
However, the six other overlapping relations were not addressed, and the testing exhaustivity was not reached for any $n$.
Oprea~\cite{oprealost} re-used Di Paolo's model to uncover IPv6 evasion and insertion opportunities on Suricata. 
If configured with Extreme Performance Tuning guidelines~\cite{septun3} and running on a Linux Ubuntu 22.04 host, he showed that exploitable attacks can target Suricata when monitoring Windows 10, OpenBSD 7.4, or FreeBSD 14.0 hosts.  

As~\Cref{sec:appendices_related_works_summary} summarizes, none of the related works provide a platform that 1) exhaustively tests any $n$ chunk sequence, 2) tests a wide range of IPv4, IPv6, and TCP network stack implementations, and 3) describes the obtained reassembly policies.
Suricata and Snort NIDSes have implicitly considered that $n = 2$-based reassembly policies could be extended beyond $n=2$ by actually applying them.
However, no work has ever checked the consistency of $n = 2$ policies with more overlapping chunks. 

\section{\Pyrolyse{} design}
\label{sec:pyrolyse_design}

\begin{figure*}[t!]
    \centering
    \includegraphics[width=0.9\textwidth]{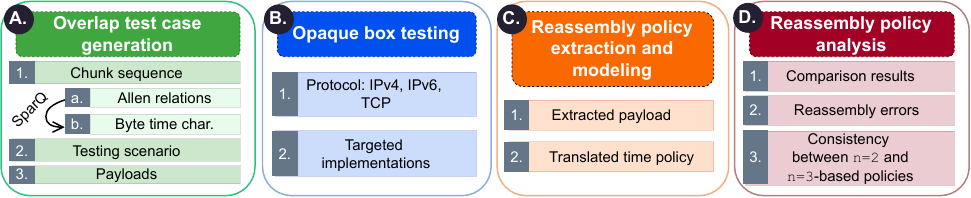}
    \caption{\label{fig:pyrolyse_test_pipeline}\Pyrolyse{} testing pipeline.
    }
\end{figure*}

\Pyrolyse{}, available at \url{https://github.com/ANSSI-FR/pyrolyse}, aims to audit various protocol implementation types when processing overlapping IPv4, IPv6, and TCP chunks. 
The targeted implementations are treated as opaque boxes to facilitate the testing setup. 
\Cref{fig:pyrolyse_test_pipeline} illustrates the tool's testing pipeline.
First, \pyrolyse{} generates exhaustive overlap test cases, thanks to Allen's spatio-temporal reasoning.
Second, it tests the IP and TCP reassemblies of various protocol implementations through a virtualizable environment (when possible) for reproducibility. 
Third, it extracts and describes the implementations' reassembly policies with a formalism easing their use. 
Fourth, \pyrolyse{} analyses the implementation policies, for example, by checking $n=2$-based policy consistency with the $n=3$ observed reassemblies.      

\subsection{Overlap test cases generation}

The test case generation step starts with the chunk sequence generation.
The chunk sequence exhaustiveness is ensured with the Allen relations, as introduced in~\Cref{sec:background_stack_reassembly_policies_n_2}. 
More precisely, \pyrolyse{} generates all the combinations of $n \choose 2$ relations. 
Using SparQ~\cite{SparQ}, \pyrolyse{} both checks if an Allen relation sequence is coherent, and, if so, computes the starting and finishing byte offsets and the time position of every chunk.
If the sequence is incoherent, the test case is invalid and thus discarded.
Then, \pyrolyse{} eventually adds real-world context (e.g., contiguous chunks located before the overlapping chunks) to the coherent chunk sequence with the testing scenario.
Finally, it populates the chunks with carefully chosen payloads to respect the upper-layer checksum correctness.

\subsubsection{Chunk sequence}

The test case's chunks offset and time characteristics are derived from the Allen relation sequence. 

\paragraph{Allen relations}

\begin{figure}[!ht]
    \centering
    \includegraphics[width=0.6\columnwidth]{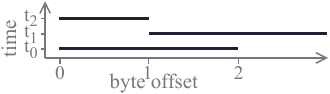}
    \caption{\label{figures/triplet_example}The ($O$, $Si$, $Mi$) triplet of chunks.}
\end{figure}

The Allen's interval algebra allows one to reason both in terms of space and time.
In our case, the space is actually the byte offset.
The 13 Allen's relations are used to generate \emph{test cases}.
Since a relation links two chunks, we define a \emph{test case for $n$ chunks} as being \emph{the $n \choose 2$ Allen's relations that link the $n$ chunks}.
The relations are described in~\Cref{tables/allen_relations}.
As we can see, four are non-overlapping relations: $M$, $Mi$, $B$, and $Bi$.
Thus, some test cases show no overlap at all.
We use the SparQ tool~\cite{SparQ} to ensure the coherence of the $n \choose 2$ Allen relations sequence.
\begin{itemize}
    \item \boldmath$n=2$\unboldmath: $2 \choose 2$ $= 1$ Allen's relation describes the link between two chunks, and the 13 relations are coherent test cases for $n=2$ chunks, as discussed in~\Cref{sec:background_stack_reassembly_policies}.
    \item \boldmath{$n=3$}\unboldmath: Since ${3 \choose 2} = 3$, three Allen relations describe all the $n = 3$ chunk test cases.
    We represent the triplet of chunks from~\Cref{figures/triplet_example} with the notation ($p_{01}$, $p_{02}$, $p_{12}$) = ($O$, $Si$, $Mi$), where indexes refer to the chunk pairs' time position.
    We find 409 coherent and unique triplet test cases.
\end{itemize}

\paragraph{Byte time characteristics} 

SparQ \emph{quantify} operation translates the coherent Allen relation sequence into a time and data offset example of the $n$ chunks. 

\subsubsection{Testing scenarii}
\label{sec:testing_scenarii}

As mentioned in~\Cref{sec:background_stack_reassembly_policies}, some related works~\cite{shankar2003active,novak2007target,ourdimva25paper} added an extra segment before (bytewise) the test case chunks and sent it after (timewise) them, to delay the targeted implementation reassembly. 
Novak and Sturges~\cite{novak2007target} and we~\cite{ourdimva25paper} also observed different reassemblies for $n=2$ depending on whether the overlaps were tested altogether or separately. 
Therefore, the testing scenario, i.e., the context surrounding the original test case chunks described with the $n \choose 2$ Allen relations, significantly impacts the test case reassembly and, ultimately, the obtained implementation reassembly policy.  
The scenarii introduced in this part aim to capture as much of the implementation policies' complexity as possible.
We differentiate two context types: one that is \emph{protocol-agnostic} and another that is \emph{protocol-dependent}.

\paragraph{Protocol-agnostic context}


\begin{table}[t!]
    \setlength\tabcolsep{10pt}
    \centering
    \caption{\label{tables/scenarii}Protocol-agnostic scenarii.
        \textcolor{Blue}{Blue} line represents \emph{Start} chunk,
        \textcolor{Red}{red} line represents \emph{End} chunk
        and \textcolor{Black}{black} line is a placeholder for the $n$ chunks that compose the test case.
    }                                                                            

    \begin{tabular}{l  c}
    \toprule
    \textbf{Name/description}        & \textbf{Illustration} \\
    \midrule
    \makecell[l]{$s^{c}$: continuous}    & \raisebox{-.5\height}{\includegraphics[width=0.16\textwidth]{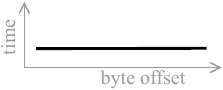}}   \\
    \hline

    \makecell[l]{$s^{sp}$:  Start precedes}     & \raisebox{-.5\height}{\includegraphics[width=0.16\textwidth]{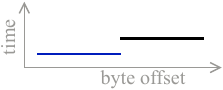}}  \\

    \hline

    \makecell[l]{$s^{sf}$: Start follows}       & \raisebox{-.5\height}{\includegraphics[width=0.16\textwidth]{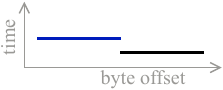}}  \\

    \hline

    \makecell[l]{$s^{ep}$: End precedes} & \raisebox{-.5\height}{\includegraphics[width=0.16\textwidth]{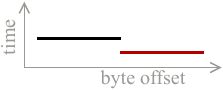}} \\

    \hline

    \makecell[l]{$s^{ef}$: End follows} & \raisebox{-.5\height}{\includegraphics[width=0.16\textwidth]{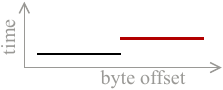}} \\

    \hline

    \makecell[l]{$s^{sf,ef}$: Start follows and \\ End follows } & \raisebox{-.5\height}{\includegraphics[width=0.16\textwidth]{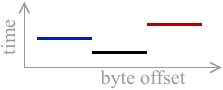}} \\

    \hline

    \makecell[l]{$s^{ef,sf}$: End follows and \\ Start follows } & \raisebox{-.5\height}{\includegraphics[width=0.16\textwidth]{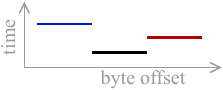}} \\
    \hline
    \makecell[l]{$s^{sp,ef}$: Start precedes and \\ End  follows } & \raisebox{-.5\height}{\includegraphics[width=0.16\textwidth]{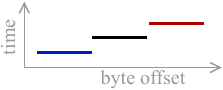}} \\
    \hline
    \makecell[l]{$s^{ep,sf}$: End precedes and \\ Start  follows } & \raisebox{-.5\height}{\includegraphics[width=0.16\textwidth]{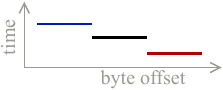}} \\
    \hline
    \makecell[l]{$s^{sp,ep}$: Start precedes and \\ End precedes } & \raisebox{-.5\height}{\includegraphics[width=0.16\textwidth]{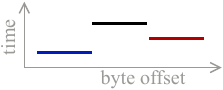}} \\
    \hline
    \makecell[l]{$s^{ep,sp}$: End precedes and \\ Start precedes } & \raisebox{-.5\height}{\includegraphics[width=0.16\textwidth]{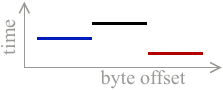}} \\

    \bottomrule
    \end{tabular}
\end{table}

An extra chunk may be added before and/or after (in terms of byte offset) the test case's chunks and sent before and/or after (in terms of time) them.
Such extra chunks are non-overlapping and contiguous with the test case's chunks.
Their addition may delay the reassembly or modify its behavior.
We refer to the (optional) added extra chunk(s) as the \emph{protocol-agnostic} context.
The related scenarii are illustrated in~\Cref{tables/scenarii} for \emph{any} test case (here represented as a black line).
In addition to the original test cases, two chunks are introduced: one before, in terms of byte offset, the original test case, named \emph{Start} and one after, named \emph{End}.
\emph{Start} and \emph{End} may timewise precede or follow the original test case or not exist at all. 
We uniquely identify 11 permutation cases by carefully crafting the scenarii names.
If both extra chunks precede or follow the original test case, the scenario name order defines which chunk was sent before. 
Furthermore, ``e''/``s" is short for ``End"/``Start" and ``p"/``f" for ``precedes"/``follows". 
The scenario $s^{ef,sf}$ thus means that the original test case chunks are sent first, then \emph{End} is sent, and finally, \emph{Start} is sent.

\paragraph{Protocol-dependent context}

Protocol semantics include peculiarities that may influence the final reassembly in the presence of data overlaps.
Specifically, we identify one peculiarity for IP protocols that comes from the More Fragment (MF) bit in the header field. 
The MF bit should be unset for the last, i.e., rightmost, fragment according to the RFC~\cite{rfc791,deering2017rfc}.
However, it is ambiguous if several rightmost fragments overlap in the context of overlapping chunks.
For example, if two or more fragments finish at the same byte offset but start at a different one, which fragment(s) must have bit MF unset?
The final fragment reassembly may change depending on the chosen strategy regarding MF bit unsetting.
In the $n \leq 3$ testing context, we introduce two sets of strategies: one in which we consider one (i.e., oldest, newest, middle) or multiple (i.e., oldest/newest, oldest/middle, middle/newest, all) \emph{rightmost finishing} chunks and,
another in which we do the same for the \emph{rightmost starting} chunks. 
We name the scenarii with the first letters of every sub-strategy. 
For instance, $of$ means that the \emph{oldest} rightmost \emph{finishing} fragment has the MF bit unset.

\paragraph{Association between protocol-agnostic and protocol-dependent parts}

Test scenarii comprise a protocol-agnostic part and an optional protocol-dependent part.
We thus suffix test scenario names with the protocol-dependent part when necessary and possible: e.g., $s^{c}_{nf}$ for IP protocols or $s^{c}$ for TCP.

The IP-related MF bit ambiguity only applies for scenarii whose protocol-agnostic context does not contain \emph{End} extra chunk (i.e., $s^{c}$, $s^{sp}$, and $s^{sf}$).
Indeed, for those with \emph{End} chunk, this extra chunk has bit MF unset.
Since TCP does not have a protocol-dependent part\footnote{We actually tested to acknowledge the target's data 1) as soon as the target sends a response and 2) only once all the test case's segments have been sent. Since we did not observe any reassembly change across the tested targets, we do not consider any TCP-dependent part.}, the TCP scenarii correspond to the one described in~\Cref{tables/scenarii}.
In total, there are 42 IP and 11 TCP scenarios, representing 10,362 and 4,642 test cases, respectively.  
Note that some IP test cases are the same in some IP-dependent scenarii (e.g., if a test case has one unique rightmost finishing fragment, this tested fragment sequence is the same in any $s_{af}$, $s_{of}$, and $s_{nf}$-related scenario).   

\subsubsection{Chunk payload population with checksum-impactless patterns}
\label{sec:pyrolyse_chunk_patterns}

We use ICMP Echo messages (resp. TCP Echo service) on top of IP to obtain IPv4 and IPv6 (resp. TCP) reassembly policies, as discussed in~\Cref{sec:opaque_box_testing}.
The IP fragments must convey the ICMP header, which contains a \emph{checksum} field. 
The ICMP checksum is computed with the 2-byte one's complement of the one's complement of the ICMP header and data. 
Besides using its header and data, ICMPv6 also uses an IP pseudo-header to compute its checksum.   
This pseudo-header is constructed with the source and destination IP addresses, the protocol, and the ICMPv6 length.

We wish to distinguish between the test case overlapping portions and that the checksum is valid no matter the implementation reassembly\footnote{For a given test case, we expect to observe varying reconstructed payload lengths for IP scenarii that test overlaps on the rightmost finishing or starting fragment(s) (i.e., the scenarii with different MF bit unsetting strategies). More generally, these patterns are introduced to prevent the checksum from impacting the implementations' reassembly. For example, this property enabled the discovery of the OpenBSD early response error described in~\Cref{sec:uc_reassembly_bugs_reassembly_logic_early_response}.}.
We thus introduce \emph{unique 8-byte-long patterns that do not impact the IP upper-layer checksum}. 
The first six bytes of a pattern correspond to its unique ID (three bytes for the chunk ID and three bytes for the pattern offset), while the last two bytes are the IP upper-layer checksum correction bytes. 
See~\Cref{sec:appendices_checksum_patterns} for more details. 

\subsection{Opaque box testing}
\label{sec:opaque_box_testing}

\Pyrolyse{} is an opaque box IPv4, IPv6, and TCP protocol testing framework.

\subsubsection{Upper-layer services}
\label{sec:upper_layer_services}

Test case reassemblies are obtained with the ICMP/ICMPv6 Echo service for IPv4/IPv6 and the TCP Echo service through port 7 for TCP. 
For the IP testing, the fragment sequences of all $n \leq 3$ test cases are first stored within PCAP files, which are then replayed with the Tcpreplay tool~\cite{tcpreplay} on the targeted implementations. 
TCP, which is a stateful protocol, requires initiating a new connection for every test case before sending the related segment sequence.
\Pyrolyse{} implements this TCP logic and enables a host to communicate with a target from the connection initialization phase to the termination, with a FIN handshake or an RST packet.
The tool acknowledges any echoed data that the targeted implementation sends.
Finally, tcpdump~\cite{tcpdump} captures IP and TCP host communications.

\subsubsection{Implementation targets}
\label{sec:implementation_targets}

\Pyrolyse{} supports testing network and midpoint stacks, such as NIDSes.  

\paragraph{Network stacks}
\label{sec:network_stacks}

Network stack testing is performed on Base/Target machine pairs. 
All the scripts are launched from the Base machines.
The targeted stacks cannot be tested similarly; we thus use three different setups:
\begin{itemize}
    \item [i.] The OSes are Vagrant/Virtualbox boxes publicly available from the Vagrant Cloud~\cite{vagrantcloud}.
    The vagrant provisions enable ICMP Echo messages and the TCP Echo service if necessary. 
    \item [ii.] We add embedded/IoT, unikernel, and DPDK-compatible stacks inside Debian boxes since these stacks are not readily available on the Vagrant Cloud. 
    In the target Vagrant provisions, the stack is compiled and bound to a TUN/TAP interface. 
    A TCP Echo server is deployed on every stack to perform the corresponding tests.
    \item [iii.] The tested NICs use proprietary hardware and network stacks that provide full TCP stack offloading. They are thus physically tested.
\end{itemize}
In all the cases, the testing outputs are PCAP files. 
Note that on OS, embedded/IoT, unikernel, and DPDK-compatible hosts, NIC offloading is disabled not to alter the test cases reaching the targeted machine, as we previously did~\cite{ourdimva25paper}.

\paragraph{NIDS midpoint stacks}
\label{sec:nids_stacks}

The NIDS official Docker container is used if it exists; otherwise, the NIDS is tested locally after compilation.  
Signature files are created for all the NIDSes, in which one signature entry is associated with precisely one of the patterns, 
and the matching buffer is the upper-layer service. 
The matching direction is from client to server. 
\Pyrolyse{} generates PCAP files for IP testing but does not synthesize TCP test cases with session initialization and termination. 
The PCAP files obtained from the TCP testing of one network stack are used to test the NIDSes. 
The NIDS testing outputs are log files. 

\subsection{Reassembly policy extraction and modeling}
\label{sec:reassembly_policy_extraction}


\begin{table}[t!]
    \setlength\tabcolsep{3pt}
    \centering

    \caption{\label{tables/policies} Description of the time policies used to describe implementation reassembly.
    $n$ is the overlapping chunk number.}        
    
    \begin{tabular}{L{0.03\columnwidth} L{0.18\columnwidth} L{0.12\columnwidth} L{0.17\columnwidth} L{0.37\columnwidth}} 
    \toprule
    \boldmath$n$         & \textbf{Overlap type} & \textbf{Notation} & \textbf{Time policy} & \textbf{Policy description}  \\
    \midrule
    \multirow{4}{*}{2} 
    & \multirow{4}{*}{Single pair} & \multirow{4}{*}{$TP_{pairs}$}
                       & old & old data is kept \\
    \cline{4-5}
    &                & & new & new data is kept \\
    \cline{4-5}
    &                & & ignores & test case chunks are ignored \\
    \hline
    \multirow{12}{*}{3}
    & \multirow{5}{*}{Residual pair} & \multirow{5}{*}{$TP_{rp,t}$}
                       & none                             & not applicable because the considered chunk pair does not overlap \\
    \cline{4-5}
    &                & & old, new, ignores                & same as above \\
    \cline{4-5}
    &                & & partialIgnore                    & only the pair overlapping data is ignored \\
    \cline{2-5}
    & \multirow{5}{*}{Triple} & \multirow{5}{*}{$TP_{t,t}$}
                       & none                             & not applicable because there is no triple overlap \\
    \cline{4-5}
    &                & & old, new, ignores                & same as above \\
    \cline{4-5}
    &                & & middle                           & data in the middle chunk is kept \\
    
    \bottomrule
    \end{tabular}
\end{table}

To obtain reassembly policies, \pyrolyse{} first extracts the test case reconstructed payload from the PCAP or log files, depending on the targeted implementation.
Then, it translates the implementation reassemblies into a time policy, in which the test case reassemblies are described regarding the temporal aspect of the preferred chunk data. 
\Cref{tables/policies} introduces this formalism to ease the analysis and use of the reassembly policies. 
The $n=2$ pair time policies from $TP_{pairs}$ introduced in~\cite{ourdimva25paper} is extended with the $n=3$ triplet time policy set $TP_{triplets}$, that is equal to $TP_{t,t} \times TP_{rp,t}^3$. 

\begin{figure}[t!]
    \centering
    \includegraphics[width = 0.7\columnwidth]{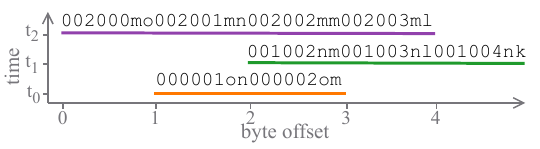}
    \caption{\label{figures/triplet}The triplet ($O$, $D$, $Oi$) representation populated with the IPv4 checksum-impactless patterns. 
    }
\end{figure}

\Cref{figures/triplet} illustrates the ($O$, $D$, $Oi$) triplet test case.
If an OS reassembles with \textcolor{Purple}{002000mo} \textcolor{Orange}{000001on 000002om} \textcolor{Green}{001003nl 001004nk} payload, the time policy is thus: $tp_t = old$ and ($tp_{rp_{01},t}$, $tp_{rp_{02},t}$, $tp_{rp_{12},t}$) = ($none$, $old$, $old$).
$tp_{rp_{01},t}$ is $none$ since the entire overlap of the \textcolor{Orange}{orange} and \textcolor{Green}{green} chunks does not overflow the boundaries of the triple overlap.

\subsection{Reassembly policy analysis}
\label{sec:reassembly_policy_analysis}

\Pyrolyse{} performs three analysis types:
\begin{itemize}
    \item The first compares reassembly policies across testing scenarii and protocol implementations to identify what part of the testing context leads to different reassemblies and to qualify and quantify implementation policy diversity. 
    \item The second one identifies the reassembly errors. 
    \Cref{sec:pyrolyse_chunk_patterns} testing payload patterns allow us to find patterns in the test case reassembled payloads that are not correctly located. 
    \item The last analysis consists of checking time policy consistencies between $n=2$ and $n=3$ test cases with different methods (that will be introduced in the corresponding~\Cref{sec:uc_reassembly_policy_generalization_temptative}) to determine whether implementation reassembly policies are generalizable to any $n$. 
\end{itemize}

\section{Use case 1: exhaustive reassembly policies for $n\leq3$ test cases and context chunks}
\label{sec:uc_pair_triplet_reassembly_policies}

The first use case for \pyrolyse{} is obtaining an IPv4, IPv6, or TCP protocol implementation reassembly policy and describing it with the~\Cref{sec:reassembly_policy_extraction} formalism. 
In this section, we give a high-level overview regarding the policies of the following implementations: \emph{OSes} (Windows, the Linux-based Debian, FreeBSD, NetBSD, OpenBSD, Solaris), \emph{embedded/IoT} (lwIP, uIP, picoTCP, smoltcp), \emph{unikernel} (mirage-tcpip), \emph{DPDK-compatible} (Seastar), and \emph{NICs} (Chelsio TOE T520-CR, Xilinx Onload 9.0.1) stacks and \emph{NIDSes} (Suricata, Snort, Zeek).
For all the targeted protocol implementations, we test the latest stable version. 
In addition, we test a range of versions from 10 years to the latest for the general-purpose OSes. 
All the tested OS versions that exhibit the very same IP and TCP reassembly policies are counted as a unique implementation in the following. 
See~\Cref{sec:appendices_target_implem} for more details on the targeted implementations. 
Full reassembly policies can be found at~\cite{pyrolysetool}.
This section first describes the implementations' reassembly policy diversity. 
Then, it characterizes the impact of the multiple testing scenarii on the test case reassembly. 

\subsection{Reassembly diversity across implementations}
\label{sec:reasembly_evolution_across_implementations}

\begin{figure}[htp]
    \centering
    \subfloat[IPv4]{\includegraphics[width=\columnwidth]{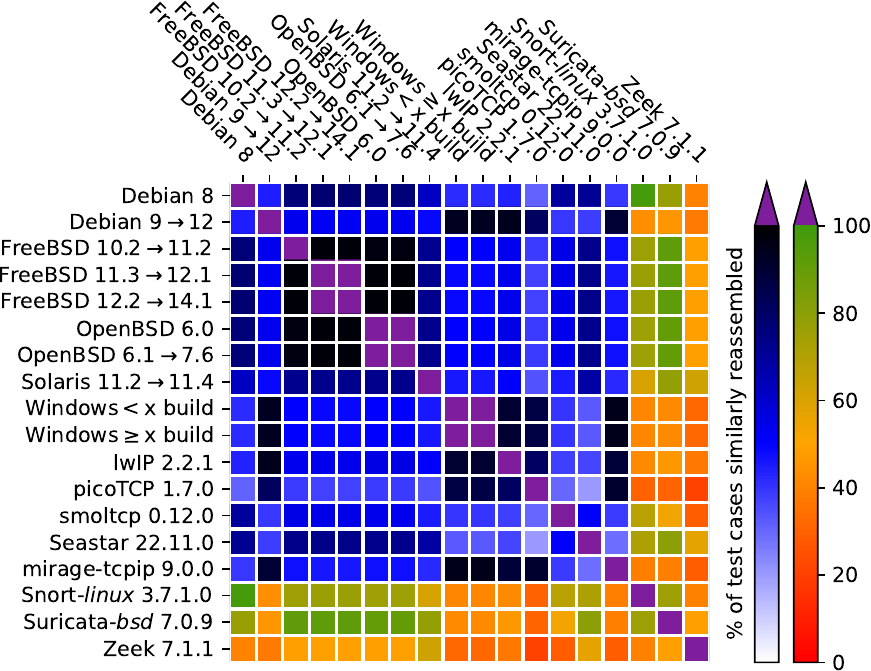}} \\
    \subfloat[IPv6]{\includegraphics[width=\columnwidth]{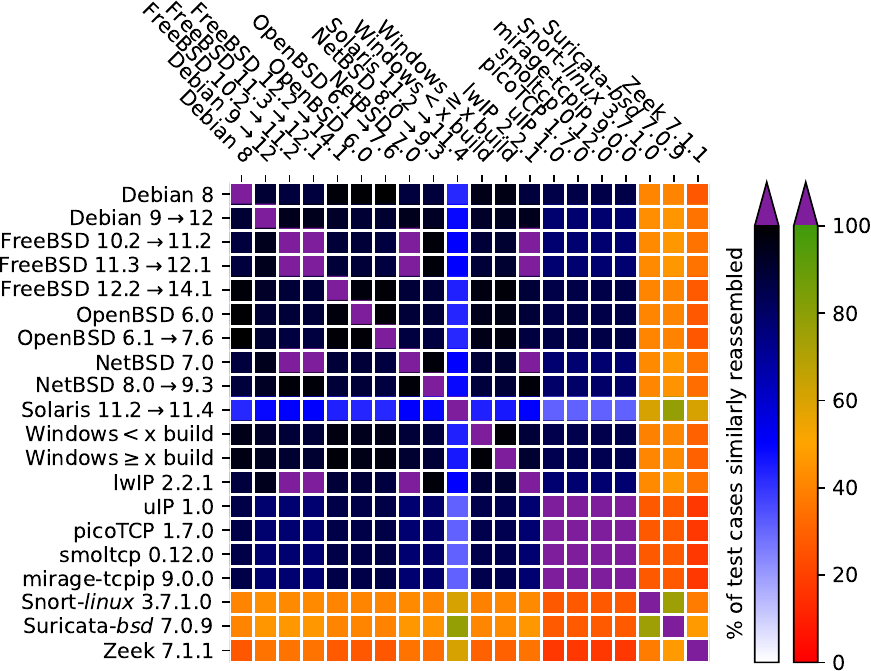}} \\
    \subfloat[TCP]{\includegraphics[width=\columnwidth]{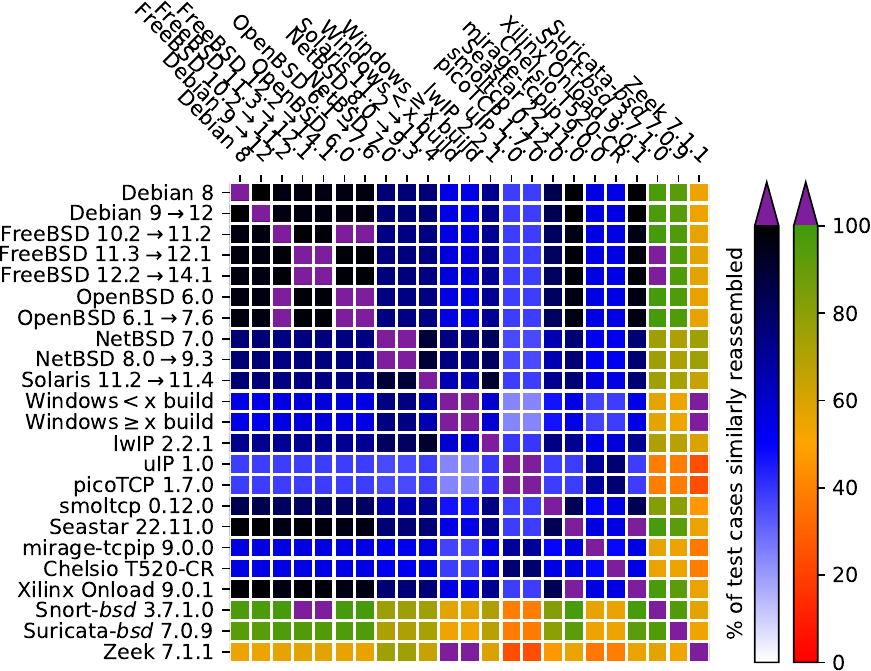}}
    \caption{\label{fig/heatmaps} IPv4, IPv6, and TCP reassembly similarities across tested implementations. 
    \textcolor{Purple}{Purple} cells encodes perfect reassembly similarities.
    \textcolor{Green}{Green}, \textcolor{Orange}{orange}, and \textcolor{Red}{red} cells encodes similarities between the NIDSes with their default reassembly policies and the other implementations. 
    \textbf{x build} corresponds to the 2022/10/25 build of the Windows OSes.  
    }
  \end{figure}

Overall, we observe 15 IPv4, 14 IPv6, and 14 TCP distinct behaviors for 18 to 23 protocol implementations.  
\Cref{fig/heatmaps} illustrates this reassembly diversity. 
Each cell reports the similarity between pairs of protocol implementations. 
For IPv4 and IPv6 (resp. TCP), the reassemblies of the 10,362 (resp. 4,642) test cases are compared. 
Similarity scores are computed as the percentage of test cases that exhibit the same time policy.
As the purple and dark squares reveal, IPv6 and TCP tend to be reassembled more similarly across implementations than IPv4.

\subsubsection{Reassembly evolution across OS versions}
\label{sec:reasembly_evolution_across_os_versions}

Five out of the six tested OS families have modified at least one of their IPv4, IPv6, or TCP reassembly policies within the last 10 years. 
In particular, the Linux-based Debian OSes changed the IPv4, IPv6, and TCP policies between Debian 8 and Debian 9. 
The most important changes impacted the IPv4 policy since only 44\% of the test cases are reassembled the same between these two Debian versions. 
IPv6 is similarly reassembled for 89\% of the test cases and 99\% for TCP. 
Interestingly, none of the TCP $n=2$ test cases are reassembled differently between versions 8 and 9. 
All the FreeBSD policies have also evolved. For IPv4 and TCP (resp. IPv6), the change occurred between versions 11.2 and 11.3 (resp. 12.1 and 12.2). 
The new policies share 98\% IPv4, 88\% IPv6, and 98\% TCP similarities with the previous ones.
Finally, Windows, OpenBSD, and NetBSD have changed their IPv6 policies slightly, with less than a 2\% difference. 
The versions introducing the change are the 2022/10/25 Windows builds\footnote{We tested this evolution for Windows Server 2019, Windows 10 (22H2), and Windows 11 (21H2) boxes.}, OpenBSD 6.1, and an unidentified NetBSD version between 7.0 and 8.0\footnote{We did not find Vagrant boxes between these two NetBSD versions.}.
The latest Windows OSes now exhibit the same reassembly behavior when processing IPv4 and IPv6 overlapping or non-overlapping fragments, which is also true for Solaris and the latest Debian OSes. 

\subsubsection{Reassembly consistency between NIDSes and network stacks}
\label{sec:reasembly_consistency_nids_and_other}

Suricata, Snort, and Zeek NIDSes reassemblies were specifically tested to measure their consistency with the stacks they are likely to monitor. 
Snort and Suricata (which allow users to configure the reassembly policies) were tested with their default policies, which are always \emph{bsd}, except for Snort and IP protocols where the \emph{linux} policy is used\footnote{
As mentioned in~\cite{ourdimva25paper}, the lack of automated configuration and the potentially high number of monitored machines make it very costly to configure a NIDS in a real-world deployment.
Therefore, we suppose the situation closest to reality uses the default NIDS reassembly policies.
}.

Only Zeek reassembles all the TCP test cases consistently with Windows OSes, always favoring the oldest data segment. 
Snort and Suricata never reassemble entirely consistently the IP test cases with any OS, including with the OS family their default policy tries to reproduce.
Since the \emph{linux} policy has evolved, Snort IPv4 inconsistencies even reach 66\% with the latest Debian OSes. 
IPv6 test cases, that Snort and Suricata both reassemble in the same way as IPv4, are never reassembled with more than 77\% similarities for Suricata and 60\% for Snort across OSes. 
Finally, Snort and Suricata reassemble TCP test cases with 0 to 4\%, 2 to 4\%, and 25 to 29\% inconsistencies with FreeBSD, OpenBSD, and NetBSD OSes, respectively. 
\emph{Every reported NIDS inconsistency can lead to an insertion or evasion attack}. 

\subsubsection{Interesting similarities and dissimilarities}
\label{sec:interesting_similarities}

IPv6 implementations reassemble the test cases quite similarly. 
This can be explained for two reasons. 
First, most of the embedded stacks do not reassemble IPv6 fragments at all.
Second, all the tested OSes but Solaris \emph{ignore} about 80\% of the test cases. 
The strong IPv4 similarity between the latest Windows and Debian OSes is also due to the many ignored test cases. 
This observation is valid for both OSes as they both exhibit identical IPv4 and IPv6 policies. 
Surprisingly, FreeBSD and OpenBSD OSes, which had the same TCP policy until FreeBSD version 11.2, have a TCP reassembly policy closer to Debian OSes than NetBSD.
Corollary, NetBSD and Solaris OSes also have close TCP reassembly policies.
BSD OSes, thus, do not have a unified TCP reassembly policy anymore.
Furthermore, mirage-tcpip, picoTCP, and uIP do not reassemble data segments that overlap with already received ones. 
Most of the time, the last two stacks and Chelsio T520-CR\footnote{The Chelsio team replied that this behavior concerns T5 ASICs and that T6 ASICs should resolve the problem, but we did not test them.} do not reassemble out-of-order segments either.
Finally, Seastar and Xilinx Onload reassemble TCP test cases similarly, and they have TCP policies close to Debian, FreeBSD, and OpenBSD OSes.

\begin{figure*}[!t]
    \centering
    \subfloat[IPv4]{\includegraphics[scale=0.36]{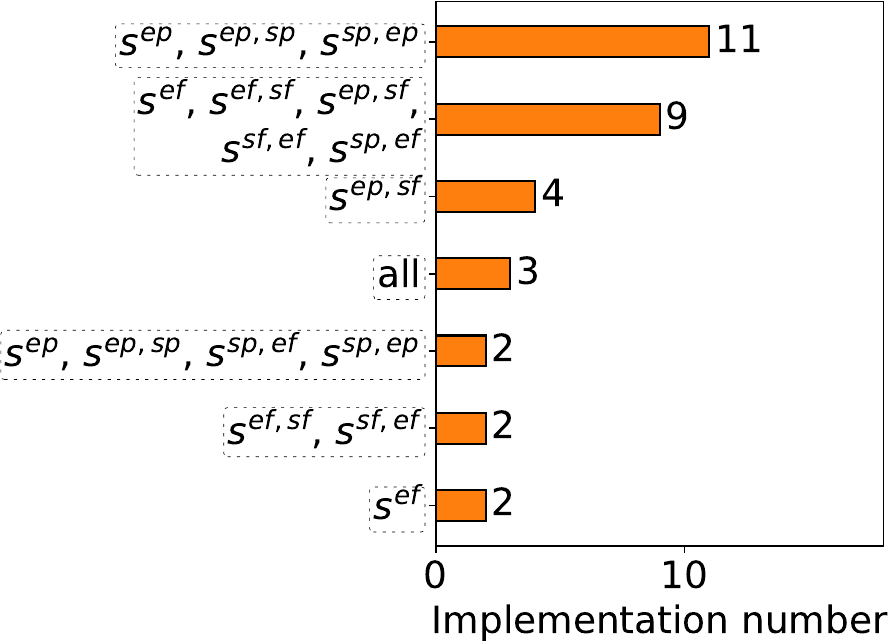}
    \label{fig/scenarii_group_pas_ipv4}}
    \hspace{0.5cm}
    \subfloat[IPv6]{%
        \includegraphics[scale=0.36]{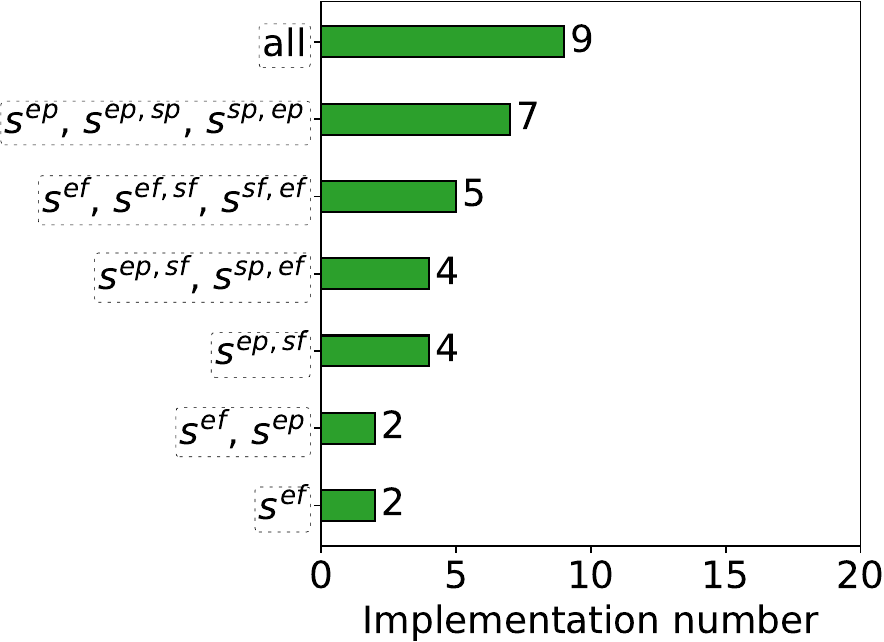}
        \label{fig/scenarii_group_pas_ipv6}    
    }
    \hspace{0.5cm}
    \subfloat[TCP]{%
        \includegraphics[scale=0.36]{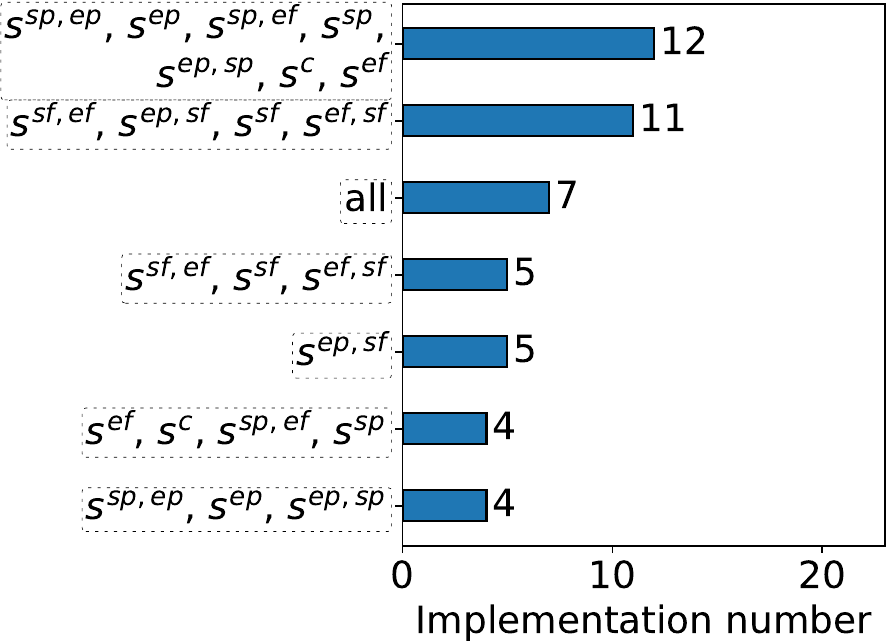}
        \label{fig/scenarii_group_pas_tcp}
    }
    \caption{\label{fig/scenarii_group_pas}Top 7 protocol-agnostic scenario groups that are similarly reassembled across implementations. The groups are sorted by the implementation number.
    }
  \end{figure*}

\subsection{Reassembly diversity across testing scenarii}

For a given implementation, \Pyrolyse{} groups the testing scenarii showing the same time policies.

\subsubsection{Protocol-agnostic scenarii}
\label{sec:uc_1_pas}

The first notable result is that $s^{ep,sp}$/$s^{sp,ep}$ and $s^{sf,ef}$/$s^{ef,sf}$ pairs are always in the same groups for any IPv4, IPv6, or TCP implementation. 
We cannot find other such pairs for the protocol-agnostic scenarii, and therefore, all the other introduced scenarii enrich and extend the reassembly policies.
\Cref{fig/scenarii_group_pas} reports the top 7 scenario groups across implementations and protocols.

\paragraph{IP}

We observe quite a lot of diversity across implementations with 14 IPv4 scenario groups for a total of eight (protocol-agnostic) scenarii in~\Cref{fig/scenarii_group_pas_ipv4}.
Eleven implementations reassemble all scenarii with $End$ preceding the test case chunks similarly, except $s^{ep,sf}$. 
This means that the hole left by $Start$ impacts these implementation reassemblies, as the top two scenario groups reveal.
Three implementations, namely Zeek, lwIP, and picoTCP, reassemble all their testing scenarii identically. 
Zeek always favors old data chunks, no matter the context. 
PicoTCP always ignores the test cases that contain only IPv4 fragments\footnote{Some $s^{c}$ test cases are actually structured in such a way that a unique chunk is enough to reassemble a complete payload.}.
The defragmentation process may thus not be present in this implementation. 

IPv6 implementations also behave in several ways, as 13 scenario groups are observed. 
Nine implementations reassemble test cases similarly across testing scenarii as shown in~\Cref{fig/scenarii_group_pas_ipv6}.
Zeek, once again, favors old data, and four (picoTCP, uIP, mirage-tcpip, and smoltcp) ignore all test cases.
The four others (FreeBSD 10.2 to 12.1, NetBSD 7.0, and lwIP) only reconstruct test cases that exhibit no overlap at all or overlaps, but, in this case, any newest overlapping fragment is entirely included in the oldest ones.
We observe the same most common scenario groups as IPv4, which contain scenarii with the $End$ preceding chunk. 
However, as for IPv4, some implementations have unique scenario groups; thus, we cannot draw any general conclusions.

\paragraph{TCP}

Eleven scenarii in total are tested. 
We observe less diversity than for IP protocols since the implementations only show seven groups of testing scenarii, as shown in~\Cref{fig/scenarii_group_pas_tcp}. 
The first two top groups show that a $Start$ segment that follows the test case sequence (thus introducing a data hole on the left) often makes the reassembly distinct from the other scenarii. 
This behavior is not surprising since TCP is a stateful protocol, and thus, data is pushed to the upper layer as long as there is no data hole on the left of the received segments. 
However, five implementations behave differently between $s^{sf}$, $s^{ef,sf}$, and $s^{sf,ef}$ on the one hand and $s^{ep,sf}$ on the other.  
As observed with IPv4, introducing multiple (at least two) data holes that are filled throughout the test case segment sequence impacts some implementations. 
Finally, Windows OSes, FreeBSD 11.3 to 14.1, Zeek, Snort, and mirage-tcpip exhibit a unique scenario group, meaning that an implementation reassembles all the test cases the same across the scenarii, with Windows OSes and Zeek notably always preferring the oldest chunk data.

\subsubsection{Protocol-dependent scenarii}

Eighty-five groups of IPv4 protocol-dependent scenarii are observed across the tested implementations. 
Apart from different versions of the same implementation, no stack has precisely the same groups of scenarii. 
The $s^{c}$, $s^{sf}$, and $s^{sp}$-like scenarii are never reassembled similarly for any implementation. 
This is expected for $s^{sf}$, but it is slightly more surprising for the $s^{c}$ and $s^{sp}$ scenarii because the reassembly conditions\footnote{For a given IPID, IP reassembly conditions are met whenever the corresponding fragment sequence exhibits no data hole and a fragment with the MF bit unset has been received.} can here be met earlier, i.e., before all the test case chunks have been received.
For the $s^{c}$-like scenarii, some test case chunks are actually complete datagrams, which explain the reassembly difference with $s^{sp}$. 
Regarding the discrepancy between unsetting the rightmost \emph{starting} and the rightmost \emph{finishing} chunk(s), we observe some similarities for the $s^{sf}$-like scenarii for Zeek, Seastar, OpenBSD, FreeBSD, lwIP, and mirage stacks. 
However, for every OS, there are always at least one couple of scenarii, e.g., $s^{sf}_{of}$/$s^{sf}_{os}$ for OpenBSD, that are reassembled differently.
More surprisingly, except Zeek and Seastar, implementations sometimes reassemble in a different way $s_{af}$ (resp. $s_{as}$) and $s_{of}$ (resp. $s_{os}$) test cases. 
It probably means these implementations drop the oldest fragments if multiple fragments have the MF bit unset. 

Slightly fewer, i.e., 83, IPv6 scenario groups are observed. 
Even if four implementations have the same protocol-dependent scenario group (since they ignore all the test cases), the IPv6 policy and, with them, the scenario groups of five OS families have evolved. 
The other IPv6 implementations behave the same as IPv4. 

\subsubsection{Impact of testing scenarii for configurable NIDSes}

As we've just discussed, the testing scenario considerably impacts the implementation reassembly, and the effect differs depending on the implementation. 
Since the diverse testing scenarii introduced in this document have never been addressed in any related work, it is unsurprising that the configurable NIDSes Suricata and Snort do not behave consistently with most OSes. 
Indeed, Suricata does not exhibit the same groups of IP and TCP scenarii as BSD OSes, nor does Snort with Debian OSes for IP or OpenBSD and NetBSD for TCP. 
The only exception is Snort with FreeBSD, which exhibits the same TCP scenario groups.
The diverse observed implementation behaviors across scenarii must be transcribed inside the NIDSes if they intend to reassemble consistently with them.  

\begin{takeaway}{Takeaway}
The 42 IP and 11 TCP scenarii and $n\leq3$ test cases uncovered diverse implementation reassembly policies (fully available at~\cite{pyrolysetool}).
They have increased from 5 IP and 6 TCP policies~\cite{novak2005target,atlasis2012attacking,novak2007target} to 15 IPv4, 14 IPv6, and 14 TCP stack behaviors.
\Pyrolyse{} provides an automated way to extract such reassembly time policies and ease testing the reassembly consistency of security equipment (such as NIDSes) with the network stack implementations.
\end{takeaway}

\section{Use case 2: reassembly errors}
\label{sec:uc_reassembly_bugs}

\renewcommand{\arraystretch}{1.1}
\begin{table}[t!]
    \setlength\tabcolsep{1pt}
    \centering
    \caption{\label{tables/bug_summary}Reassembly error summary uncovered by \pyrolyse{}. 
    \textbf{\textcolor{Orange}{!}} means that the error is partially fixed. 
    }    
    \begin{tabular}{L{0.08\textwidth} L{0.065\textwidth} L{0.09\textwidth} L{0.135\textwidth} C{0.04\textwidth} C{0.05\textwidth}}
    \toprule
    \textbf{Type} & \textbf{Impacted protocol} & \textbf{Impacted implem.} & \textbf{Possible security issue} & \textbf{CIA triad} & \textbf{Fixed?} \\ 
    \midrule
    \multirow{4}{0.08\textwidth}{Reassembly logic} & \multirow{2}{0.065\textwidth}{IPv4/IPv6} & Suricata    & \multirow{2}{0.135\textwidth}{pattern-matching bypass} &  & \textbf{\textcolor{Orange}{!}} \\ 
    &&  Snort &  &  &\redxmark \\
    \cline{2-6}
    & IPv4 &  OpenBSD  & payload shortening & I &\greencmark \\ 
    \cline{2-6}
    & TCP & Suricata  & pattern-matching bypass &  &\textbf{\textcolor{Orange}{!}} \\ 
    \hline
    \multirow{3}{0.08\textwidth}{Connection termination} & \multirow{3}{*}{TCP} & OpenBSD  & \multirow{3}{0.135\textwidth}{resource exhaustion/ DoS}& \multirow{3}{*}{A} &\greencmark  \\
    & & lwIP  & &  &\redxmark \\
    & & mirage-tcpip  &&  &\redxmark \\
    \hline
    Log display & TCP & Suricata  &  security investigation hindrance &  &\greencmark \\
    \bottomrule
\end{tabular}
                                                                        
\end{table}

This section details the reassembly errors \pyrolyse{} discovered.
After manual investigation, we identify three categories of errors: reassembly logic, connection termination, and log display. 
One OS (OpenBSD), two NIDSes (Suricata and Snort), and two other stacks (lwIP and mirage-tcpip) are impacted. 
In total, we reported eight bugs to the implementation developers; half are entirely or partially fixed.
\Cref{tables/bug_summary} summarizes the encountered errors.

\subsection{Reassembly logic}
\label{sec:uc_reassembly_bugs_reassembly_logic}

The reassembly logic category groups the errors that directly affect the reconstructed payloads.
The observed bugs lead to data holes or data located after a data hole in the reassembled payload or an early response from the targeted implementation. 
The hole-related issues impacted Suricata and Snort NIDSes. 
This error is critical since it damages the NIDS's ability to perform its key functionality, namely, malicious pattern-matching. 
Suricata partially fixed the problem for IP and TCP, but Snort did not.
The early response was observed while testing the OpenBSD IPv4 stack.

\subsubsection{Data holes in the reassembled payload}
\label{sec:uc_reassembly_bugs_reassembly_logic_data_hole_1}

\begin{figure}[t!]
    \centering
    \includegraphics[width = 0.8\columnwidth]{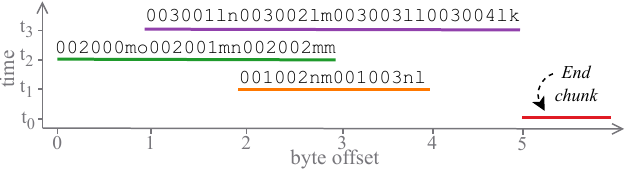}
    \caption{\label{images/test_408_peoep}The ($Oi$, $O$, $D$) test case representation in a $s^{ep}$ testing context.
    }
\end{figure}

Snort and Suricata allow users to configure the reassembly policy according to the monitored OS IP addresses. 
The reassembly error impacts each NIDS's reassembly policies differently.

\paragraph{Suricata}

The error is present in 93 to 801 test cases, impacts about half of the IP scenarii, and concerns all Suricata reassembly policies (including the default one \emph{bsd}).
One impacted test case is the triplet illustrated in~\Cref{images/test_408_peoep}.
With the \emph{default} policy, the reconstructed datagram is 
\textcolor{Green}{002000mo} \textcolor{Green}{002001mn} \textcolor{Orange}{001002nm} \textcolor{Orange}{001003nl} \textbf{........}.
Since Suricata raises no alert related to the \textcolor{Purple}{003004lk} pattern, it is blind between offsets 4 and 5, making the pattern-matching functionality inoperable. 
After some investigation, we concluded that an incorrect reconstructed packet length calculation and an early exit along the fragment queue processing caused the behavior. 
This reassembly error was assigned the CVE 2024-32867~\cite{cve_ours} and is partially fixed from Suricata version 7.0.5.

\paragraph{Snort}

The error concerns all reassembly policies (including the default one \emph{linux}) for 26 to 211 triplet test cases. 
The behavior is observed for some triplet test cases containing a third fragment that overlaps the previous two, such as ($Oi$, $O$, $D$) of~\Cref{images/test_408_peoep}.
Snort's reassembly is 
\textcolor{Green}{002000mo} \textcolor{Green}{002001mn} \textcolor{Green}{002002mm} \textcolor{Orange}{001003nl} \textbf{........} 
with the \emph{bsd} policy. 
The NIDS uses a variable $slide$ to track the amount of data to remove from the third fragment. 
However, after the $p_{12}$ overlap resolution (i.e., the choice to ignore or to reassemble with old or new data), the $slide$'s value is overwritten by the $p_{02}$ overlap resolution, which subsequently impacts the final datagram reconstruction.  

\subsubsection{Reassembly with data located after a hole}
\label{sec:uc_reassembly_bugs_reassembly_logic_data_hole_2}

\begin{figure}[t!]
    \centering
    \includegraphics[width = 0.45\columnwidth]{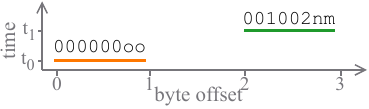}
    \caption{\label{images/test_1_pep}The ($B$) test case representation in a $s^{c}$ testing context.
    }
\end{figure}

For any TCP scenario and test cases \emph{with a hole}, Suricata reassembles with some data located after (in terms of byte offset) the data hole.
Since a test case with a hole is characterized with one (resp. two or three) Allen's before-like relation, i.e., $B$ or $Bi$, for $n = 2$ (resp. $n = 3$), there are two over 13 (resp. 72 over 409) of a scenario's test cases that are impacted, which represent a total of 814 test cases. 
\Cref{images/test_1_pep} illustrates one impacted test case.
The observed Suricata reassembly is \textcolor{Orange}{000000oo} \textcolor{Green}{001002nm} from the \emph{payload\_printable} field of the eve.log file. 
Suricata also raised an alert for both \textcolor{Orange}{000000oo} and \textcolor{Green}{001002nm} patterns. 
The NIDS can then be fooled if a malicious data segment fills the data hole from 1 to 2.  
This issue is partially fixed from Suricata version 7.0.7.

\subsubsection{Early response}
\label{sec:uc_reassembly_bugs_reassembly_logic_early_response}

OpenBSD responded to the fragment sequence for 14 triplet cases tested with $s^{ef}$, $s^{sp,ef}$, and $s^{sf,ef}$ scenarii, whereas it did not receive all the fragments.
More precisely, OpenBSD responded before processing the fragment with the MF bit unset.
It resulted in the truncation of the original datagram, i.e., the reassembled datagram length was less than it should have been. 
This is how \pyrolyse{} was able to discover the error.
The latter was introduced in OpenBSD's code while implementing DoS protection in pf\footnote{Pf is the OpenBSD solution for packet filtering. It reassembles overlapping IPv4 fragments.} fragment handling on 2018/09/04.
The code avoided traversing "the list of fragment entries to check whether the pf(4) reassembly is complete" by using a data hole counter.  
The bug is now patched~\cite{openbsdpatchip}.

\subsection{Connection termination}
\label{sec:uc_reassembly_bugs_connection_termination}

A similar error was observed during the TCP testing for three tested implementations: OpenBSD, mirage-tcpip, and lwIP. 
This bug always led to the non-termination of some test case-related TCP connections. 
The impacted test cases were, however, different across the implementations. 

On the one hand, OpenBSD did not reset the TCP connection if there had previously been a data hole during that connection. 
Several sequence numbers were tested to reset it without success. 
This reassembly error was observed in all the testing scenarii and is now patched~\cite{openbsdpatchtcp}.

On the other hand, mirage-tcpip 9.0.0 and lwIP 2.2.1 echoed some data but not the maximum possible amount for the test cases in question. 
We believe they considered the reset packet sequence number invalid because it was located after all the data was sent (including the overlapping segments). 
Since they did not echo the maximum possible amount of data, the initiator SND.NXT internal variable was probably not the same as the RCV.NXT value in mirage-tcpip and lwIP. 

\subsection{Log display}
\label{sec:uc_reassembly_bugs_log_display}

Suricata reassemblies were obtained through log files using two methods: the signatures and the reconstructed payload-related log field. 
Suricata's log payload field contained duplicate payload patterns for about 72\% to 84\% of the TCP test cases, depending on the scenario. 
This manifested the same way as a reassembly logic error, since some patterns from the reconstructed payload were not correctly located. 
However, after an unsuccessful attempt to raise an alert on the duplicated reconstructed payload log field, we classified the bug as a log display error. 
After some investigation, we discovered that the signatures and the payload field do not use the same TCP reassembly buffers, which explains the divergent behaviors across extraction payload methods. 
The bug is not critical because we do not observe the pattern duplication in the buffer used for the pattern-matching.
However, it can confuse the NIDS security investigators, who may think of a false positive alert and, ultimately, misclassify a malicious security event.
Suricata version 7.0.7 patched this reassembly error.

\begin{takeaway}{Takeaway}
    \Pyrolyse{} found eight reassembly errors while testing the 23 targeted implementations with the $n\leq3$-related 42 IP and 11 TCP testing scenarii. 
    Three reassembly logic bugs impact Suricata and Snort NIDSes and are critical because they can lead to a malicious pattern-matching functionality bypass.
    Developers wholly or partially fixed half of them. 
\end{takeaway}

\section{Use case 3: reassembly inconsistency between $n=2$ and $n=3$ test cases and custom reassembly algorithm testing}
\label{sec:uc_reassembly_policy_generalization_temptative}

This section aims to link the observed $n = 2$ and $n = 3$ implementation reassemblies. 
The configurable Suricata and Snort NIDSes currently apply $n = 2$ pair policies to any $n$ overlapping chunk sequence. 
When an overlapping chunk arrives, it is compared to already received ones, one by one, and the pair policies are applied. 
First, the section shows that the current NIDS approach is incorrect since one testing scenario does not exhibit any inconsistency between some OSes and NIDSes for $n=2$, but many for $n=3$.
Therefore, $n = 2$ policies cannot be used alone to reassemble any $n$ sequence inside the NIDSes.
Second, it identifies some root causes for the observed inconsistencies. 
Finally, considering these root causes, it explores the extension of $n = 2$ pair time policies (as described in~\Cref{sec:reassembly_policy_extraction}) with manually implemented reassembly algorithms, as a way for the NIDSes to reassemble any $n$ sequence.

\renewcommand{\arraystretch}{0.9}
\begin{table}[t!]
    \setlength\tabcolsep{2.5pt}
    \centering
    \small

    \caption{\label{table:triplet_tcp_nids_inconsistencies}TCP test case inconsistency number across couple latest OS/NIDS reassembly policy, for the $s^{sf}$ testing scenario.}
    \begin{tabular}{l l c c}
        \toprule

        \multirow{2}{*}{\textbf{OS}} & \multirow{2}{*}{\textbf{NIDS policy}} & \multicolumn{2}{c}{\textbf{Number of inconsistent test cases}}                         \\
        \cline{3-4}
                                       &          & \boldmath$n=2$ (/13)                                           & \boldmath$n=3$ (/409) \\
        \midrule
        \multirow{2}{*}{Windows 11} & Suricata-\emph{first}                  & \textcolor{Green}{0} & \textcolor{Red}{12}                                         \\
         & Snort-\emph{first}                  & \textcolor{Green}{0} & \textcolor{Green}{0}                                         \\
        \hline
        \multirow{2}{*}{Debian 12} & Suricata-\emph{linux}                   & \textcolor{Green}{0} & \textcolor{Red}{21}                                         \\
         & Snort-\emph{linux}                   & \textcolor{Green}{0} & \textcolor{Red}{15}                                         \\
        \hline
        \multirow{2}{*}{Solaris 11.4} &Suricata-\emph{solaris}                 & \textcolor{Red}{1}  & \textcolor{Red}{40}                                          \\
         &Snort-\emph{solaris}                 & \textcolor{Red}{1}  & \textcolor{Red}{36}                                          \\
        \hline
        \multirow{2}{*}{OpenBSD 7.6} &                                                                                                            Suricata-\emph{bsd}                   & \textcolor{Green}{0} & \textcolor{Red}{30}                                       \\
        &                                                                                                            Snort-\emph{bsd}                   & \textcolor{Green}{0} & \textcolor{Red}{42}                                       \\
        \bottomrule
    \end{tabular}

\end{table}

\subsection{Suricata and Snort's inconsistencies with OSes for $n=3$ overlapping segments}
\label{sec:uc_reassembly_policy_generalization_temptative_ids_inconsistencies}

Snort and Suricata users can associate host IP addresses with a set of implemented reassembly policies. 
As a preliminary approach to verify if $n=3$ test case reassemblies can be deduced from $n=2$-based policies, we compare the reassembly consistency of Suricata and Snort with some OSes. 
Since these NIDS-implemented policies are based on Novak and Sturges's work~\cite{novak2007target}, we use our related TCP scenario $s^{sf}$.
\Cref{table:triplet_tcp_nids_inconsistencies} shows that both Suricata and Snort reassemble all $n=2$ test cases similarly with the latest Windows, Debian, and OpenBSD OSes. 
However, the two NIDSes reassemble differently with these OSes some $n=3$ test cases, except Snort with Windows.
Therefore, the reassembly consistency between a NIDS and an OS for $n=2$ chunk sequences does not necessarily imply any kind of reassembly consistency for $n \geq 3$ chunk sequences. 
Overall, the implicit direct application of $n=2$-based implementation policies when $n \geq 3$ inside the NIDSes is incorrect (except for Snort and Windows).

\subsection{Inconsistency between n=2 and n=3 reassemblies: observations and root cause example}

We here first report an observation from the $n \leq 3$ observed policies regarding to $n=2$ and $n=3$ inconsistencies.
Then, we detail one root cause for some of these inconsistencies.

\subsubsection{Observations from the $n \leq 3$ observed policies}

\begin{figure}[t!]
    \centering
    \includegraphics[width = 0.7\columnwidth]{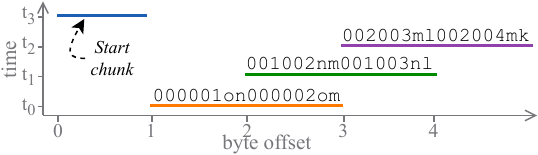}
    \caption{\label{images/test_440_peosf}The ($O$, $M$, $O$) test case representation in a $s^{sf}$ testing context.
    }
\end{figure}

As discussed in~\Cref{sec:reasembly_evolution_across_os_versions}, some OSes updated their reassembly policies within the last 10 years.
These updates impact $n=3$ test case reassemblies without necessarily affecting the $n=2$ ones, as reported for Linux's TCP stack. 
For example, Debian 8 and Debian 9 to 12 reassemble the ($O$) pair favoring the oldest data in a $s^{sf}$ context. 
Yet, Debian 8 and Debian 9 to 12 reconstruct differently the ($O$, $M$, $O$) test case, which is illustrated in~\Cref{images/test_440_peosf}, 
i.e., \textcolor{Orange}{000001on} \textcolor{Orange}{000002om} \textcolor{Green}{001003nl} \textcolor{Purple}{002004mk} vs \textcolor{Orange}{000001on} \textcolor{Orange}{000002om} \textcolor{Purple}{002003ml} \textcolor{Purple}{002004mk}.
This separated evolution proves that $n=3$ implementation policies cannot be derived --- at least solely --- from the $n=2$ testing.

\subsubsection{Impact analysis of the chunk merging mechanism}
\label{sec:chunk_merging}

One interesting function involved in the Linux 6.1 (i.e., our Debian 12's kernel) TCP reassembly process is $try\_to\_coalesce()$~\cite{linuxmergingfunction}.
This function seems to merge data segments, as the source code function description mentions "before queuing skb\footnote{Short for socket buffer.}, try to merge them to reduce overall memory use and queue lengths, if cost is small".

\begin{figure}[!ht]
    \centering
    \includegraphics[width = 0.6\columnwidth]{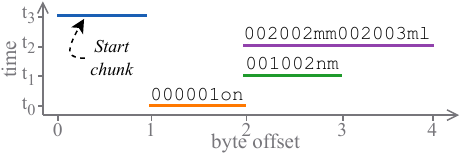}
    \caption{\label{images/test_368_peosf}The ($M$, $M$, $S$) test case representation in a $s^{sf}$ testing context.
    }
\end{figure}

A protocol implementation that merges chunks, as Linux 6.1 seems to do, reassembles differently from what is expected for some triplet test cases. 
\Cref{images/test_368_peosf} illustrates such a test case.

Let's consider the Debian 12 TCP reassembly policy in which the $S$ overlap is reassembled favoring the newest data while $O$ is with the oldest.
\begin{itemize}
 \item \ul{Expected reassembly (i.e., without merging) from $n=2$ policy}:
 the \textcolor{Green}{green} and \textcolor{Purple}{purple} segments overlaps with $S$ relation, newest data is preferred.
 The resulting reassembled data is \textcolor{Orange}{000001on} \textcolor{Purple}{002002mm 002003ml}.
 \item \ul{Non-expected reassembly (i.e., with merging) from $n=2$ policy}:
 the \textcolor{Orange}{orange} and \textcolor{Green}{green} segments are contiguous, they are thus merged.
 \textcolor{Orange}{Orange}-\textcolor{Green}{green} and \textcolor{Purple}{purple} segments' corresponding Allen relation is now $O$, which is reassembled favoring the oldest segment data.
 The resulting reassembly is \textcolor{Orange}{000001on} \textcolor{Green}{001002nm} \textcolor{Purple}{002003ml}.
\end{itemize}
This reassembly mechanism may modify the chunk configuration, the associated Allen relations, and therefore, the applied time policies.

\subsection{Reassembling $n = 3$ test cases with custom algorithms}
\label{sec:uc_3_custom_algo}

One way for the NIDSes with configurable policies to reassemble consistently the $n \leq 3$ triplet test cases with the stacks is to apply the $n \leq 3$ time policies in the same direct way as the current $n = 2$-based policies.
This can be done through 1) multiple conditional branches or 2) a time policy hashmap look-up.
However, 1) is very tedious and error-prone for $n \leq 3$ due to the large test case number and the implementation policy diversity (as described in~\Cref{sec:uc_pair_triplet_reassembly_policies}).
Doing 2) requires resolving the overlaps once all the chunks have been received and computing the corresponding Allen relation sequence. 
Furthermore, these solutions do not ensure the NIDSes are consistent for any $n$.   
Considering the $n = 2$ and $n = 3$ inconsistency causes introduced in the last part, another approach for the NIDSes would be to extend reassembly policies using custom-implemented algorithms that leverage observed reassembly mechanisms such as the one described in~\Cref{sec:chunk_merging}.
If these algorithms are consistent with $n = 3$ cases, they might also be correct for $n > 3$. 
Additional tests would, however, need to be performed to verify this aspect.
We implement some algorithms inside \pyrolyse{} and attempt to reproduce the $n = 3$ implementation behaviors.  
First, we describe these algorithm principles and then analyze the consistency of the reassembly between the algorithms and the implementations.

\subsubsection{Principle}

As previously discussed, protocol implementation reassembly policies based on $n=2$ testing are not easily extensible to any $n$.
We hypothesize that addressing more testing scenarii than state-of-the-art and $n=3$ test cases brings more reassembly algorithm parts into play.
It gives a more complete but also more complex picture of reassembly policies. 
We characterize reassembly algorithms using three main components:
\begin{itemize}
    \item [(i)] the function(s) or code that we consider to directly impact the overlap resolution, i.e., the choice to ignore or to reassemble with old or new data.
    Suppose some overlaps are not addressed in the code (i.e., no action is explicitly taken for the overlap resolution). 
    In that case, it likely means that the overlap is undetected, and consequently, the implementation would overwrite data on the overlapping portions.
          Both cases lie in this item: they belong to \emph{the overlap resolution component}.
    \item [(ii)] the mechanisms that indirectly impact the overlap resolution in the scope of our testing. 
          Such mechanisms may, for instance, be present for performance reasons (e.g., chunk merging~\cite{linuxmergingfunction}) or due to protocol specificities (e.g., MF bit information tracking for IP).
          They belong to the \emph{reassembly mechanisms with side effect component}.
    \item [(iii)] the mechanisms that have no direct or indirect impact on the reconstructed payload in the scope of our testing.
          Such mechanisms belong to the \emph{reassembly mechanisms with no side effect component}.
\end{itemize}

The intuition is that the implementations perform the overlap resolution (i) by pairs of two chunks. 
Thus, $n=2$ testing alone can capture (i) behavior.
However, other (ii)-like mechanisms may also impact the final test case reassembly, as seen in~\Cref{sec:chunk_merging}.
These mechanisms explain, at least to some extent, the Suricata and Snort's inconsistent reassemblies for $n=3$ reported in~\Cref{sec:uc_reassembly_policy_generalization_temptative_ids_inconsistencies}.
We thus hypothesize that by considering $n=2$ reassembly policies coupled with some indirectly implicated reassembly mechanisms, the NIDSes may be able to reassemble consistently with the implementations. 
We implemented custom reassembly algorithms to verify this point. 
The algorithms are further detailed in~\Cref{sec:appendices_custom_reassembly_algorithm_test}.   

\begin{table}[t!]
    \setlength\tabcolsep{1.5pt}
    \centering

    \caption{\label{tables/reassembly_algo_inconsistency_summary} Reassembly consistency of the 3,376 IP and 4,642 TCP ($n\leq3$) protocol-agnostic test cases across implementations.
    }   
    \vspace{-2mm}
    \begin{tabular}{L{0.06\textwidth} L{0.04\textwidth} L{0.12\textwidth} L{0.04\textwidth} L{0.19\textwidth}}
    \toprule
    \multirow{2}{*}{\textbf{Protocol}} & \multicolumn{4}{c}{\textbf{Consistency between...}} \\
    \cline{2-5}
    & \multicolumn{2}{l}{\textbf{\makecell{...\boldmath$n=2$ and \boldmath$n=3$\\residual pairs inside\\implementations (baseline)}}} & \multicolumn{2}{l}{\textbf{\makecell{...implementations and custom\\reassembly algorithms \boldmath$n=3$\\time policies}}} \\
    \cline{2-5}
    & \textbf{\#} & \textbf{Implem. names} & \textbf{\#} & \textbf{Implem. names} \\

    \midrule
    IPv4 & 1/18 &  picoTCP  & 6/18 & Windows OSes, lwIP, picoTCP, Seastar, mirage-tcpip \\ 
    \hline
    IPv6 & 4/20 & picoTCP, uIP, mirage-tcpip, smoltcp & 10/20 & Windows OSes, FreeBSD 10.2 to 11.2, FreeBSD 11.3 to 12.1, NetBSD 7.0, lwIP, picoTCP, mirage-tcpip, uIP, smoltcp \\ 
    \hline
    TCP & 2/23 & Windows OSes & 9/23 & Windows, OpenBSD, and FreeBSD OSes, smoltcp, Seastar \\ 
    \bottomrule
    \end{tabular}
        \vspace{-5mm}
\end{table}

\subsubsection{Triplet test case consistency analysis}

The consistency performance of custom reassembly algorithms is exposed in~\Cref{tables/reassembly_algo_inconsistency_summary}. 
The second column lists implementations whose time policy of residual pairs inside triplets (i.e., we ignore triple overlap as defined in~\Cref{sec:reassembly_policy_extraction}) is consistent with $n=2$-based time policies. 
The rightmost column reports the implementations in which at least one reassembly algorithm successfully reproduces its $n=3$ behavior.
The second-column consistency check method is more limited than the third-column since it does not consider the triple overlap (if there is any). 
It is thus unusable in practice, but we add it as a baseline.
The implementations that exhibit the best consistency between $n=2$ pairs and residual pairs are some of those that only showed one scenario group in~\Cref{sec:uc_1_pas}. 
They either always favor the oldest chunks or ignore all test cases.
The third column shows that the custom algorithms explain more implementation triplet policies than the baseline (while also trying to predict triple overlaps).
These implementation behaviors for $n\leq3$ can thus all be abstracted and are easily explained. 
For example, Seastar merges any overlapping or non-overlapping IPv4 fragments and TCP segments after the overlap resolution. 
Another example is NetBSD 7.0, lwIP, FreeBSD 10.2 to 12.1, which drop the newest fragment when they ignore an IPv6 $n=2$ test case (which explains the related observation in~\Cref{sec:uc_1_pas}).
See the related mechanism explanation in~\Cref{sec:appendices_reassembly_mechanisms_with_side_effect}.
However, more than 10 stack policies cannot wholly be inferred for each protocol.

\begin{takeaway}{Takeaway}
Most implementations show inconsistencies between $n=2$ pair and $n=3$ triplet test cases. 
This is mainly due to some reassembly mechanisms (for example, introduced for performance reasons) that have a side effect on the applied policy.
Consequently, NIDSes must not use $n=2$-based time policies alone to reassemble any $n$ overlapping chunk sequences. 
\Pyrolyse{} implements some custom reassembly algorithms composed of sets of observed reassembly mechanisms. 
They achieve reassembly consistency for only 25 of the 61 protocol stacks.
\end{takeaway}

\section{Discussion}

\subsection{Implementation reassembly policy compliance with RFC}

IPv4 and TCP RFC~\cite{rfc791,rfc9293} do not specify the behavior implementations must adopt in the presence of fragment or segment data overlaps. 
However, in the TCP receive window definition, the specification says that "[...] segments overlapping the range RCV.NXT to RCV.NXT + RCV.WND - 1 carry acceptable data [...]. Segments containing sequence numbers entirely outside this range are considered duplicates or injection attacks and discarded".
Consequently, even if the specification is unclear, we believe that accepting overlapping data segments is the most RFC-compliant behavior. 
PicoTCP, uIP, and mirage-tcpip do not reassemble overlapping segments.

Differently, IPv6~\cite{deering2017rfc} states that implementations must drop the entire fragment flow if any of the bytes overlap. 
As reported in~\Cref{sec:reasembly_evolution_across_implementations}, picoTCP, uIP, mirage-tcpip, and smoltcp stacks ignored all the test cases; they are thus perfectly compliant with the standard. 
Even if the other IPv6 implementations but Solaris \emph{ignore} about 80\% of the test cases (see~\Cref{sec:uc_pair_triplet_reassembly_policies}), we still observe overlapping test cases that are reassembled, which is non-compliant with the RFC but consistent with the related works~\cite{atlasis2012attacking,di2023new,lin2024research}. 

\subsection{NIDS challenges to address more than n = 2 overlapping chunks}

Addressing various testing scenarii and exhaustive sequences of three overlapping chunks showed that implementation policies are much more diverse than previously thought. 
Pair ($n=2$) and triplet ($n=3$) test case reassemblies are often inconsistent because reassembly mechanisms have side effects on triplet test case structure and thus reassembly.
Therefore, $n=2$-based policies cannot be used alone as they currently are inside configurable NIDSes.
The generic reassembly algorithms we implemented abstract stack behaviors, but only achieve consistency with 25 of the 61 protocol stacks.
The remaining algorithm inconsistencies may be due to some other non-implemented reassembly mechanisms and/or unwanted introduced behaviors that are very implementation-specific.
One could investigate the former by implementing and testing mechanisms that have not yet been addressed. 
For 44 out of the 61 protocol stacks, the custom algorithms improve $n = 2$ and $n = 3$ consistency and explainability compared to the baseline method. 
This is promising but not satisfying from a NIDS perspective, since a unique inconsistency can lead to the NIDS bypass.
Moreover, while $n > 3$ is not tested in this paper, we strongly suspect that similar stack reassembly inconsistencies between $n = 2$ and $n > 3$ would be observed if tested.
Consequently, we recommend that the NIDSes that still wish to offer policy configurability use the $n=2$ pair time policies in any case and raise an alert as soon as processing a $n > 2$ overlapping chunk sequence. 
The currently implemented $n=2$ time policies inside the NIDSes should be enriched with all the testing scenarii addressed in this paper, since they reproduce real-world context. 
Furthermore, the research community should revisit overlapping chunk occurrence and structure in the wild since the last work on this topic was from 2008~\cite{john2008detection}. 
This would help NIDS developers assess if alerts on $n > 2$ overlapping chunk sequences would produce false positive alerts.

\section{Ethical considerations}

\subsection{Responsible Disclosure}

\subsubsection*{Reassembly errors}

Every reassembly error described in~\Cref{sec:uc_reassembly_bugs} was reported to the impacted implementation developers at least one month before this paper submission:  
\begin{itemize}
    \item OpenBSD fixed the IPv4 and TCP issues. 
    \item Suricata completely fixed the TCP log display and partially addressed the reassembly logic-related errors. 
    We informed Suricata developers that some bugs could still be observed with their fix, but did not receive any answer. 
    The CVE-2024-32867~\cite{cve_ours} was assigned to the IP error. 
    \item mirage-tcpip is currently analyzing the reassembly error. 
    \item Snort and lwIP did not respond to our disclosure.
\end{itemize} 
The detailed process is described in~\Cref{sec:responsible_disclosure_process}. 

\subsubsection*{Implementation reassembly policies}

To our knowledge, only Suricata and Snort NIDSes have configurable reassembly policies.
Both IDSes based them on Novak and Sturges's works~\cite{novak2005target,novak2007target}, which date back to 2005 and 2007.
Since this paper contains both an extension of prior OS reassembly policies~\cite{novak2005target,novak2007target,ourdimva25paper} and new aspects (e.g., testing scenarii), we gradually communicated Suricata and Snort's results so they can address them progressively.
We sent the first report and related artifacts (i.e., implementation policy descriptions) to Suricata and Snort in January 2024 and the last one in March 2025.
We sent the same reports to Zeek, another widely deployed NIDS that previously offered the policy configurability feature.
We did not receive any response.

\subsection{Censorship Systems}

Improving NIDS security and performance has the side effect of enhancing censorship systems.
Some works showed that data overlaps could circumvent CSes until 2017~\cite{khattak2013towards,wang2017your}. 
But subsequent works noticed that this approach became ineffective~\cite{bock2019geneva,wang2020symtcp}.
Thus, our results should not affect the censorship elusion techniques currently used.
Nonetheless, even if this strategy is in use to circumvent censorship systems, we consider that improving defense capabilities outweighs the negative impacts on censorship elusion techniques. 

\section{Conclusion}

In this paper, we described \pyrolyse{}, which we used to test the behavior of a wide range of IPv4, IPv6, and TCP protocol implementations when processing overlapping chunk sequences in various testing scenarii. 
The tool identified eight reassembly errors during the testing campaign, including reassembly logic, connection termination, and NIDS log display errors. 
One was assigned a CVE~\cite{cve_ours}, and five are completely or partially fixed. 
The obtained OS, embedded/IoT, unikernel, NIC, or DPDK-compatible stack reassembly policies with \pyrolyse{} revealed many distinct behaviors, as 15 IPv4, 14 IPv6, and 14 TCP policies were observed over the 23 tested implementations. 
Suricata, Snort, and Zeek NIDSes reassemble inconsistently with the other tested protocol stacks, except the TCP tests for Zeek with Windows OSes and Snort with FreeBSD 11.3 to 14.1.
Since we did not find a way to explain the observed $n=3$ overlapping test case reassemblies for all the protocol implementations, we state that the obtained policies cannot be extended from $n=2$ pairs to any $n$ overlapping IP fragment or TCP segment sequences.
We provide precise related recommendations for the NIDSes.

\section*{Acknowledgment}

This work was supported by a grant from the French National Cybersecurity Agency (ANSSI).
We thank Olivier Levillain and Gregory Blanc for their help in chunk sequence modeling, and Zhengyu Zu for his work in OS TCP testing.
We also thank CERT-FR that guided us through the responsible disclosure process as well as stack and NIDS developers for their constructive answers and corrections: Jason Ish and Victor Julien for Suricata, Alexander Bluhm for OpenBSD, Nicolas Tsiftes for uIP, Anil Madhavapeddy for mirage-tcpip, Thibaut Vandervelden for smoltcp, Tom Reu for Chelsio, and Michael Tuexen, Timo Voelker, Gleb Smirnoff, Rodney W. Grimes, and Olivier Cochard for FreeBSD.
We are grateful to Kosek et al.~\cite{kosek2020must} for their help with lwIP and uIP testing.
Finally, we thank the reviewers for their insightful comments and our shepherd, Andrea Oliveri, for his guidance.

\bibliographystyle{IEEEtran}
\bibliography{references} 

\begin{thebibliography}{10}
\providecommand{\url}[1]{#1}
\csname url@samestyle\endcsname
\providecommand{\newblock}{\relax}
\providecommand{\bibinfo}[2]{#2}
\providecommand{\BIBentrySTDinterwordspacing}{\spaceskip=0pt\relax}
\providecommand{\BIBentryALTinterwordstretchfactor}{4}
\providecommand{\BIBentryALTinterwordspacing}{\spaceskip=\fontdimen2\font plus
\BIBentryALTinterwordstretchfactor\fontdimen3\font minus \fontdimen4\font\relax}
\providecommand{\BIBforeignlanguage}[2]{{%
\expandafter\ifx\csname l@#1\endcsname\relax
\typeout{** WARNING: IEEEtran.bst: No hyphenation pattern has been}%
\typeout{** loaded for the language `#1'. Using the pattern for}%
\typeout{** the default language instead.}%
\else
\language=\csname l@#1\endcsname
\fi
#2}}
\providecommand{\BIBdecl}{\relax}
\BIBdecl

\bibitem{cve_ours}
\BIBentryALTinterwordspacing
NIST. (2024) {CVE-2024-32867}. [Online]. Available: \url{https://nvd.nist.gov/vuln/detail/CVE-2024-32867}
\BIBentrySTDinterwordspacing

\bibitem{rfc791}
\BIBentryALTinterwordspacing
``{Internet Protocol},'' RFC 791, 1981. [Online]. Available: \url{https://www.rfc-editor.org/info/rfc791}
\BIBentrySTDinterwordspacing

\bibitem{rfc9293}
\BIBentryALTinterwordspacing
W.~Eddy, ``{Transmission Control Protocol (TCP)},'' RFC 9293, 2022. [Online]. Available: \url{https://www.rfc-editor.org/info/rfc9293}
\BIBentrySTDinterwordspacing

\bibitem{deering2017rfc}
S.~Deering and R.~Hinden, ``{RFC 8200: Internet protocol, version 6 (ipv6) specification},'' 2017.

\bibitem{ptacek1998insertion}
T.~Ptacek and T.~Newsham, ``Insertion, evasion, and denial of service: Eluding network intrusion detection,'' Secure Networks, Inc, Tech. Rep., 1998.

\bibitem{novak2005target}
J.~Novak, \emph{Target-based fragmentation reassembly}, 2005.

\bibitem{novak2007target}
J.~Novak and S.~Sturges, \emph{Target-based tcp stream reassembly}, 2007.

\bibitem{suricata}
\BIBentryALTinterwordspacing
Suricata. [Online]. Available: \url{https://suricata.io/}
\BIBentrySTDinterwordspacing

\bibitem{roesch1999snort}
M.~Roesch \emph{et~al.}, ``Snort: Lightweight intrusion detection for networks.'' 1999.

\bibitem{rpsuricata}
\BIBentryALTinterwordspacing
Suricata reassembly policies. [Online]. Available: \url{https://docs.suricata.io/en/suricata-7.0.4/configuration/suricata-yaml.html#host-os-policy}
\BIBentrySTDinterwordspacing

\bibitem{rpipsnort}
\BIBentryALTinterwordspacing
Snort ip reassembly policies. [Online]. Available: \url{https://snort.org/faq/readme-frag3}
\BIBentrySTDinterwordspacing

\bibitem{rptcpsnort}
\BIBentryALTinterwordspacing
Snort tcp reassembly policies. [Online]. Available: \url{https://snort.org/faq/readme-stream5}
\BIBentrySTDinterwordspacing

\bibitem{ourdimva25paper}
L.~Aubard, J.~Mazel, G.~Guette, and P.~Chifflier, ``{Overlapping data in network protocols: bridging OS and NIDS reassembly gap},'' in \emph{DIMVA}, 2025.

\bibitem{khattak2013towards}
S.~Khattak, M.~Javed, P.~D. Anderson, and V.~Paxson, ``Towards illuminating a censorship monitor's model to facilitate evasion,'' in \emph{FOCI}, 2013.

\bibitem{wang2017your}
Z.~Wang, Y.~Cao, Z.~Qian, C.~Song, and S.~Krishnamurthy, ``Your state is not mine: A closer look at evading stateful internet censorship,'' in \emph{ACM IMC}, 2017.

\bibitem{bock2019geneva}
K.~Bock, G.~Hughey, X.~Qiang, and D.~Levin, ``Geneva: Evolving censorship evasion strategies,'' in \emph{ACM CCS}, 2019.

\bibitem{wang2020symtcp}
Z.~Wang and S.~Zhu, ``{SymTCP: eluding stateful deep packet inspection with automated discrepancy discovery},'' in \emph{NDSS}, 2020.

\bibitem{suricatacve2}
\BIBentryALTinterwordspacing
NIST. (2021) {CVE-2021-37592}. [Online]. Available: \url{https://nvd.nist.gov/vuln/detail/CVE-2021-37592}
\BIBentrySTDinterwordspacing

\bibitem{suricatacve1}
\BIBentryALTinterwordspacing
------. (2024) {CVE-2024-55629}. [Online]. Available: \url{https://nvd.nist.gov/vuln/detail/CVE-2024-55629}
\BIBentrySTDinterwordspacing

\bibitem{amosyscve}
\BIBentryALTinterwordspacing
------. (2024) {CVE-2024-37151}. [Online]. Available: \url{https://nvd.nist.gov/vuln/detail/CVE-2024-37151}
\BIBentrySTDinterwordspacing

\bibitem{li2017lib}
F.~Li, A.~Razaghpanah, A.~M. Kakhki, A.~A. Niaki, D.~Choffnes, P.~Gill, and A.~Mislove, ``{lib.erate(n) a library for exposing (traffic-classification) rules and avoiding them efficiently},'' in \emph{ACM IMC}, 2017.

\bibitem{wang2021themis}
Z.~Wang, S.~Zhu, K.~Man, P.~Zhu, Y.~Hao, Z.~Qian, S.-V. Krishnamurthy, T.~La~Porta, and M.-J. De~Lucia, ``Themis: Ambiguity-aware network intrusion detection based on symbolic model comparison,'' in \emph{MTD}, 2021.

\bibitem{bock2020come}
K.~Bock, G.~Hughey, L.-H. Merino, T.~Arya, D.~Liscinsky, R.~Pogosian, and D.~Levin, ``Come as you are: Helping unmodified clients bypass censorship with server-side evasion,'' in \emph{ACM SIGCOMM}, 2020.

\bibitem{bock2021even}
K.~Bock, G.~Naval, K.~Reese, and D.~Levin, ``Even censors have a backup: Examining china's double https censorship middleboxes,'' in \emph{ACM SIGCOMM}, 2021.

\bibitem{atlasis2012attacking}
A.~Atlasis, ``Attacking ipv6 implementation using fragmentation,'' \emph{Black Hat}, 2012.

\bibitem{di2023new}
E.~Di~Paolo, E.~Bassetti, and A.~Spognardi, ``{A New Model for Testing IPv6 Fragment Handling},'' in \emph{ESORICS}, 2023.

\bibitem{shankar2003active}
U.~Shankar and V.~Paxson, ``Active mapping: Resisting nids evasion without altering traffic,'' in \emph{SP}, 2003.

\bibitem{oprealost}
I.-C. Oprea, ``Lost in reassembly: Exploiting ip fragmentation in computer networks,'' Master's thesis.

\bibitem{septun3}
\BIBentryALTinterwordspacing
P.~Manev and A.~Herz, ``{Suricata Extreme Performance Tuning: SepTun Mark III},'' Presented at Suricon 2024. [Online]. Available: \url{https://suricon.net/wp-content/uploads/2024/12/SuriCon2024-Peter-Manev_Andreas-Herz_Suricata-Extreme-Performance-Tuning-SepTun-Mark-III.pdf}
\BIBentrySTDinterwordspacing

\bibitem{SparQ}
\BIBentryALTinterwordspacing
Sparq tool. [Online]. Available: \url{https://github.com/dwolter/SparQ}
\BIBentrySTDinterwordspacing

\bibitem{tcpreplay}
\BIBentryALTinterwordspacing
Tcpreplay tool. [Online]. Available: \url{https://tcpreplay.appneta.com/}
\BIBentrySTDinterwordspacing

\bibitem{tcpdump}
\BIBentryALTinterwordspacing
tcpdump tool. [Online]. Available: \url{https://www.tcpdump.org}
\BIBentrySTDinterwordspacing

\bibitem{vagrantcloud}
\BIBentryALTinterwordspacing
Vagrant cloud. [Online]. Available: \url{https://app.vagrantup.com/boxes/search}
\BIBentrySTDinterwordspacing

\bibitem{pyrolysetool}
\BIBentryALTinterwordspacing
``{PYROLYSE tool and paper artifacts},'' 2025. [Online]. Available: \url{https://github.com/ANSSI-FR/pyrolyse}
\BIBentrySTDinterwordspacing

\bibitem{openbsdpatchip}
\BIBentryALTinterwordspacing
A.~Bluhm, ``{OpenBSD patch related to pf's IP reassembly error.}'' [Online]. Available: \url{https://github.com/openbsd/src/commit/9915416fe885bd67bf19209c5fe2f9bbbb7191ab}
\BIBentrySTDinterwordspacing

\bibitem{openbsdpatchtcp}
\BIBentryALTinterwordspacing
------, ``{OpenBSD patch related to pf's TCP reassembly error.}'' [Online]. Available: \url{https://github.com/openbsd/src/commit/12e4c257ea8089c7f56eb0c2c4493a0458273f4f}
\BIBentrySTDinterwordspacing

\bibitem{linuxmergingfunction}
\BIBentryALTinterwordspacing
``The try\_to\_coalesce() function source code.'' [Online]. Available: \url{https://elixir.bootlin.com/linux/v6.1/source/net/ipv4/tcp_input.c#L4653}
\BIBentrySTDinterwordspacing

\bibitem{lin2024research}
B.~Lin, L.~Zhang, Y.~Guo, H.~Zhang, and Y.~Fang, ``Research on security protection evasion mechanism based on ipv6 fragment headers,'' in \emph{IEEE LCN}, 2024.

\bibitem{john2008detection}
W.~John and T.~Olovsson, ``Detection of malicious traffic on back-bone links via packet header analysis,'' \emph{CWIS}, 2008.

\bibitem{kosek2020must}
M.~Kosek, L.~Bl{\"o}cher, J.~R{\"u}th, T.~Zimmermann, and O.~Hohlfeld, ``{MUST, SHOULD, DON'T CARE: TCP Conformance in the Wild},'' in \emph{International Conference on Passive and Active Network Measurement}.\hskip 1em plus 0.5em minus 0.4em\relax Springer, 2020, pp. 122--138.

\bibitem{song2002fragroute}
\BIBentryALTinterwordspacing
D.~Song, ``Fragroute tool,'' 2002. [Online]. Available: \url{https://www.monkey.org/~dugsong/fragroute/}
\BIBentrySTDinterwordspacing

\bibitem{intang}
\BIBentryALTinterwordspacing
``{INTANG tool},'' 2017. [Online]. Available: \url{https://github.com/seclab-ucr/INTANG}
\BIBentrySTDinterwordspacing

\bibitem{geneva}
\BIBentryALTinterwordspacing
``Geneva tool,'' 2019. [Online]. Available: \url{https://github.com/Kkevsterrr/geneva}
\BIBentrySTDinterwordspacing

\bibitem{symtcp}
\BIBentryALTinterwordspacing
``{SymTCP tool},'' 2020. [Online]. Available: \url{https://github.com/seclab-ucr/SymTCP}
\BIBentrySTDinterwordspacing

\bibitem{themis}
\BIBentryALTinterwordspacing
``Themis tool,'' 2021. [Online]. Available: \url{https://github.com/seclab-ucr/Themis}
\BIBentrySTDinterwordspacing

\bibitem{ipv6fragmentationtool}
\BIBentryALTinterwordspacing
``Ipv6-fragmentation tool,'' 2023. [Online]. Available: \url{https://github.com/netsecuritylab/ipv6-fragmentation}
\BIBentrySTDinterwordspacing

\bibitem{frageva6}
\BIBentryALTinterwordspacing
``Ipv6-fragmentation tool,'' 2023. [Online]. Available: \url{https://github.com/linbin89/FragEva6-Guard}
\BIBentrySTDinterwordspacing

\bibitem{krishnan2009rfc}
S.~Krishnan, ``{RFC 5722: Handling of Overlapping IPv6 Fragments},'' 2009.

\bibitem{opreatool}
\BIBentryALTinterwordspacing
``Oprea's tool,'' 2025. [Online]. Available: \url{https://github.com/OpreaCristian2002/reassembly-policies-5g-using-ipv6}
\BIBentrySTDinterwordspacing

\bibitem{brandt2018domain}
M.~Brandt, T.~Dai, A.~Klein, H.~Shulman, and M.~Waidner, ``Domain validation++ for mitm-resilient pki,'' in \emph{CCS}, 2018.

\bibitem{zheng2020poison}
X.~Zheng, C.~Lu, J.~Peng, Q.~Yang, D.~Zhou, B.~Liu, K.~Man, S.~Hao, H.~Duan, and Z.~Qian, ``Poison over troubled forwarders: A cache poisoning attack targeting $\{$DNS$\}$ forwarding devices,'' in \emph{{USENIX Security}}, 2020.

\bibitem{feng2022pmtud}
X.~Feng, Q.~Li, K.~Sun, K.~Xu, B.~Liu, X.~Zheng, Q.~Yang, H.~Duan, and Z.~Qian, ``{PMTUD is not panacea: Revisiting IP fragmentation attacks against TCP},'' in \emph{NDSS}, 2022.

\end{thebibliography}

\appendix

\subsection{Related works comparison}
\label{sec:appendices_related_works_summary}

\Cref{tables/state_of_the_art_summary} compares the related works and present paper contributions. 

\renewcommand{\arraystretch}{1}

\begin{table*}[t!]
    \setlength\tabcolsep{1.5pt}
    \centering
    \caption{\label{tables/state_of_the_art_summary}Summary regarding overlap-based works.
    $S^{IP}$ (resp. $S^{TCP}$) corresponds to the 42 IP (resp. 11 TCP) scenarii introduced in~\Cref{sec:testing_scenarii}. 
    \textbf{*} means that the source code is present inside the paper. 
    $(1)$ means several chunks composed the test cases but the chunks overlap only pairwise (it refers to the \emph{multiple} mode in~\cite{ourdimva25paper}). 
    $(2)$ means the method/tool should be exhaustive in theory, but is not because it would exhaust resources or require significant changes to the testing method/tool.
    $(3)$ means the authors instrumented the network target (i.e., non-opaque-box approach); thus, it requires significant effort to adapt the tool to other network targets.
    }
    \begin{tabular}{L{0.08\textwidth}C{0.04\textwidth}c ccccccccc cc}
        \toprule
        \multirow{2}{0.08\textwidth}{\textbf{Author}}                                  & \multirow{2}{*}{\textbf{Work}}                                    & \multirow{2}{*}{\textbf{Year}}  & \multirow{2}{*}{\textbf{Protocol}}            & \multirow{2}{*}{\textbf{\makecell{Testing                                                                \\scenario}}} & \multirow{2}{*}{\textbf{\makecell{\boldmath$n$\\overlaps}}} & \multirow{2}{*}{\textbf{\makecell{Test case\\exhaustivity}}} & \multicolumn{3}{c}{\textbf{Tool characteristics}} & \multirow{2}{*}{\textbf{\makecell{Reassembly policy\\format}}} & \multicolumn{2}{c}{\textbf{\makecell{Target stack}}} \\
        \cline{8-10}
        \cline{12-13}
        &                                     &   &             &  &  &  & \textbf{available} & \textbf{\makecell{target\\extensible}} & \textbf{\makecell{\boldmath$n$\\extensible}} & & \textbf{midpoint} & \textbf{network} \\
        \midrule
        Ptaceck and Newsham & \cite{ptacek1998insertion} & 1998 & IPv4/TCP & $s^c$ & 2 &  \redxmark & \greencmark~\cite{song2002fragroute} &  \redxmark &  \redxmark & natural language & NIDS & OS  \\
        \hline
        \multirow{2}{0.08\textwidth}{Shankar and Paxson} & \multirow{2}{0.04\textwidth}{\cite{shankar2003active}} & \multirow{2}{*}{2003} & IPv4 & $s^{c}$ & 2$^{(1)}$ & \redxmark & \multirow{2}{*}{\redxmark} & \multirow{2}{*}{\redxmark} &  \multirow{2}{*}{\redxmark} & \multirow{2}{*}{natural language} & & \multirow{2}{*}{\makecell{OS/router/\\printer}} \\
        \cline{4-7}
         &  &  & TCP & $s^{sf}$ & 2 & \redxmark \\
        \hline
        \multirow{2}{0.08\textwidth}{Novak and Sturges}   & \cite{novak2005target}                                & 2005    & IPv4                         & $s^{c}$           & \multirow{2}{*}{2} & \multirow{2}{*}{\greencmark} & \multirow{2}{*}{*} & \multirow{2}{*}{\redxmark} & \multirow{2}{*}{\redxmark} & \multirow{2}{*}{natural language} & & \multirow{2}{*}{OS} \\
        \cline{2-5}
                                                         & \cite{novak2007target}    & 2007 & TCP         & $s^{sf}$           & \\
        \hline
        Atlasis                                          & \cite{atlasis2012attacking}  & 2012                             & IPv6                         & \makecell{$s^{c}_{nf}$\\$s^{c}_{mnf}$}                  & 3                  & \redxmark & * & \redxmark &  \redxmark & figures & & OS                  \\
        \hline
        Khattak$\:$et$\:$al.                             & \cite{khattak2013towards}   & 2013                              & IPv4/TCP                     & ?             & 2                 & \greencmark & \redxmark & \redxmark & \redxmark  & natural language &  CS   &               \\
        \hline
        \multirow{3}{0.08\textwidth}{Wang et al.} & \multirow{3}{0.04\textwidth}{\cite{wang2017your}} & \multirow{3}{*}{2017} & IPv4 & $s^{sf}_{af}$ & 2 & \redxmark & \multirow{3}{*}{\greencmark~\cite{intang}} & \multirow{3}{*}{\greencmark} & \multirow{3}{*}{\redxmark} &  \multirow{3}{*}{natural language} & \multirow{3}{*}{CS} & \\
        \cline{4-7}
         &  &  & \multirow{2}{*}{TCP} & $s^{sf}$ & \multirow{2}{*}{2} & \multirow{2}{*}{\redxmark} &  &  &  &  &  & \\
         &  &  & & $s^{c}$ & &  &  &  &  &  &  & \\
        \hline
        \multirow{3}{0.08\textwidth}{Bock et al.} & \cite{bock2019geneva} & 2019 & \multirow{3}{*}{IPv4/TCP} & \multirow{3}{*}{-} & \multirow{3}{*}{-} & \multirow{3}{*}{\redxmark$^{(2)}$} & \multirow{3}{*}{\greencmark~\cite{geneva}} & \multirow{3}{*}{\greencmark} & \multirow{3}{*}{-} & \multirow{3}{*}{\makecell{Geneva's genetic\\building blocks\\+ natural language}} & \multirow{3}{*}{CS} & \\
        \cline{2-3}
         & \cite{bock2020come} & 2020 & &  & &  &  &  &  &  &  & \\
        \cline{2-3}
         & \cite{bock2021even} & 2021 & &  & &  &  &  &  &  &  & \\
        \hline
        \multirow{2}{0.08\textwidth}{Wang et al.} & \cite{wang2020symtcp} & 2020 & TCP & - & 2/3 & \redxmark$^{(2)}$ &  \greencmark~\cite{symtcp} & \greencmark$^{(3)}$ & \greencmark & natural language & CS/NIDS & OS$^{(3)}$ \\
        \cline{2-13}
        & \cite{wang2021themis} & 2021 & TCP & - & 3 & \redxmark$^{(2)}$ &  \greencmark~\cite{themis} & \greencmark$^{(3)}$ & \greencmark & natural language &  & OS$^{(3)}$ \\
        \hline
        \makecell[l]{Di Paolo et\\al.} & \cite{di2023new} & 2023 & IPv6 & $s^{c}_{af}$ & 2$^{(1)}$ & \redxmark & \greencmark~\cite{ipv6fragmentationtool} & \greencmark & \redxmark & nb. of target responses &  & OS \\
        \hline
        Lin et al. & \cite{lin2024research} & 2024 & IPv6  & $s^{c}$ & 2$^{(1)}$ &\redxmark & \greencmark~\cite{frageva6} & \greencmark & \redxmark & \makecell{compliance with\\RFC 5722~\cite{krishnan2009rfc}} & \makecell{NIDS/\\firewall} &  OS \\
        \hline
        Oprea & \cite{oprealost} & 2025 & IPv6 & $s^{c}_{af}$ & 2$^{(1)}$ & \redxmark & \greencmark~\cite{opreatool} & \greencmark & \redxmark & \makecell{nb. of exploitable\\overlap. sequences} & NIDS & OS \\
        \hline
        \multirow{2}{0.08\textwidth}{Us}                                      & \multirow{2}{*}{\cite{ourdimva25paper}}                                 & \multirow{2}{*}{2025}       & IPv4/IPv6                                                                                             & $s^{c}_{nf}$ & \multirow{2}{*}{2} & \multirow{2}{*}{\greencmark} & \multirow{2}{*}{\redxmark} & \multirow{2}{*}{\redxmark} & \multirow{2}{*}{\redxmark} & \multirow{2}{*}{\makecell{Allen relations\\+ pair time policies}}  & \multirow{2}{*}{NIDS} &\multirow{2}{*}{OS}\\
        \cline{4-5}
        & & & TCP & $s^{sf}$ \\
        \midrule
        \midrule
        \multirow{3}{0.08\textwidth}{\textbf{Us}}                                      & \multirow{3}{*}{\textbf{-}}                                           & \multirow{3}{*}{\textbf{-}} & \textbf{IPv4/IPv6}                                                                                             & \boldmath$S^{IP}$ & \multirow{3}{*}{\textbf{2/3}} & \multirow{3}{*}{\greencmark} & \multirow{3}{*}{\greencmark~\cite{pyrolysetool}} & \multirow{3}{*}{\greencmark} & \multirow{3}{*}{\greencmark}  & \multirow{3}{*}{\textbf{\makecell{Allen relation sequence\\+ pair/triplet time\\policies}}} & \multirow{3}{*}{\textbf{NIDS}} & \multirow{3}{*}{\textbf{\makecell{OS/IoT/\\unikernel/NIC/\\DPDK-comp.}}}\\
        \cline{4-5}
        & & & \multirow{2}{*}{\textbf{TCP}} & \multirow{2}{*}{\boldmath$S^{TCP}$} \\
        & \\
        \bottomrule
    \end{tabular}
\end{table*}

\subsection{Additional information regarding the checksum-impactless patterns}
\label{sec:appendices_checksum_patterns}

For the IP testing, the IP fragments must convey the ICMP header, which contains a \emph{checksum} field. 
The ICMP checksum is computed with the 2-byte one's complement of the one's complement of the ICMP header and data. 
Besides using its header and data, ICMPv6 also uses an IP pseudo-header to compute its checksum.   
This pseudo-header is constructed with the source and destination IP addresses, the protocol, and the ICMPv6 length.

\paragraph{Requirements}

First, the overlapping data portions must be different to distinguish which chunk data is preferred.
Second, the patterns must be 8-byte-long since this corresponds to the IP \emph{Fragment Offset} unit.   
Third, the IP upper-layer checksum must be correct no matter the final test case reassembly (i.e., no matter the chosen overlapping portion pattern and no matter the reassembled ICMP Echo Request payload length). 
As a result, we need \emph{unique 8-byte-long patterns that do no impact on the IP upper-layer checksum}.

\paragraph{Design}

\begin{figure}[!ht]
    \centering
    \includegraphics[width = 0.9\columnwidth]{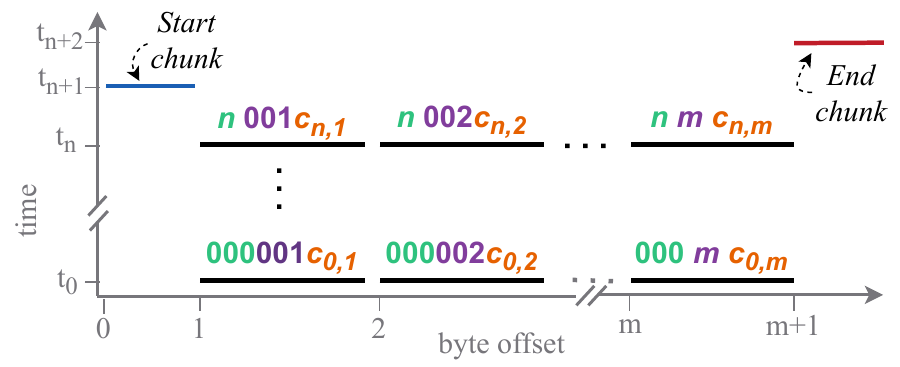}
    \caption{\label{images/invariant_checkum_patterns} The checksum-impactless patterns, in a $s^{sf,ef}$ testing scenario. 
    The first three bytes (\textcolor{Green}{green} parts) correspond to the chunk ID.
    The following three bytes (\textcolor{Purple}{purple} parts) correspond to the pattern offset.
    The last two bytes (\textcolor{Orange}{orange} parts) are the checksum correction bytes. 
    }
\end{figure}

\Cref{images/invariant_checkum_patterns} illustrates the set of chunk patterns that are generated inside \pyrolyse{}.
The first six bytes of a pattern correspond to its unique ID while the last two bytes are the IP upper-layer checksum correction bytes. 
Since the checksum is a 2-byte word sum, we can manipulate the last two bytes so that any pattern ones' complement contribution to the upper-layer checksum is null (as already outlined for UDP and TCP upper-layers in related works~\cite{brandt2018domain, zheng2020poison,feng2022pmtud}). 
Because the ICMPv6 length is also used to compute the checksum, the last IPv4 and IPv6 patterns' two bytes are different for the same pattern ID. 

\subsection{Additional information regarding the targeted implementations}
\label{sec:appendices_target_implem}

The~\Cref{tables/protocol_implementation_targets} describes the protocol implementation that are targeted in the present paper.

\renewcommand{\arraystretch}{1.3}
\begin{table*}[t!]
    \setlength\tabcolsep{3pt}
    \centering
    
    \caption{\label{tables/protocol_implementation_targets}Protocol implementation target descriptions.
    The OS reassemblies provided in the paper are those with the firewall \textit{default} state.  
    If any reassembly is different when the firewall is enabled/disabled, they are documented in~\cite{pyrolysetool}.
    }  

    \begin{tabular}{C{0.07\textwidth} L{0.1\textwidth} C{0.1\textwidth} L{0.15\textwidth} c L{0.4\textwidth}}
    \toprule
    \multicolumn{4}{c}{\textbf{Implementation}} & \multirow{2}{*}{\textbf{\makecell{Addressed\\protocol}}} & \multirow{2}{*}{\textbf{Comments}} \\ 
    \cline{1-4}
    \textbf{Type} & \textbf{Name} & \textbf{Version} & \textbf{\makecell{Additional\\version info}} \\ 
    \midrule
    \multirow{13}{*}{OS} & \multirow{2}{0.065\textwidth}{Windows} & \multirow{2}{0.1\textwidth}{Server 2019/ 10/11} & $<$2022/10/25 build  & \multirow{2}{*}{IPv4/IPv6/TCP} & \multirow{2}{0.4\textwidth}{Windows' firewall is enabled by default, and we did not observe any reassembly difference if disabled.} \\
    \cline{4-4}
    &  & &  $\geq$2022/10/25 build &   &  \\
    \cline{2-6}
    & \multirow{3}{0.065\textwidth}{Debian} & 8  & Linux 3.16.0-9 & \multirow{3}{*}{IPv4/IPv6/TCP} & \multirow{3}{*}{-} \\
    \cline{3-4}
    &  & 9  & Linux 4.9.0-12  &  \\
    \cline{3-4}
    &  & 12 & Linux 6.1.0-37  &  \\
    \cline{2-6}
    & \multirow{6}{0.065\textwidth}{FreeBSD} & 10.2 & RELEASE-p8 &\multirow{6}{*}{IPv4/IPv6/TCP} & \multirow{6}{0.4\textwidth}{The parameters related to IP reassembly are dynamically computed based on the FreeBSD host's RAM. In RAM scenarii lower than 1GB, we observe IP reassembly inconsistencies across the runs. We do not oberve any reassembly difference with pf, ipfw, and ipfilter firewalls enabled/disabled.} \\
    \cline{3-4}
    &  & 11.2 & RELEASE-p10 & & \\
    \cline{3-4}
    &  & 11.3 & RELEASE-p9 & & \\
    \cline{3-4}
    &  & 12.1 & RELEASE-p10 & & \\
    \cline{3-4}
    &  & 12.2 & RELEASE-p6 & & \\
    \cline{3-4}
    &  & 14.1 & RELEASE & & \\
    \cline{2-6}
    & \multirow{4}{0.065\textwidth}{OpenBSD} & \multirow{2}{*}{6.0} & \multirow{2}{*}{GENERIC.MP\#2319} & \multirow{6}{*}{IPv4/IPv6/TCP} & \multirow{6}{0.4\textwidth}{OpenBSD's firewall is enabled by default, and we observe IPv4 and IPv6 reassembly difference if disabled. Note that the IP reassembly error (\Cref{sec:uc_reassembly_bugs}) is fixed from OpenBSD version 7.7 or by applying the related patch in previous versions (e.g., \url{https://ftp.openbsd.org/pub/OpenBSD/patches/7.6/common/007_pffrag.patch.sig}). The TCP fix will appear in version 7.8. In~\Cref{sec:uc_pair_triplet_reassembly_policies,sec:uc_reassembly_policy_generalization_temptative}, we report OpenBSD 7.6 reassemblies AFTER the IP and TCP fixes.} \\
    & \\
    \cline{3-4}
    &  & \multirow{2}{*}{6.1}   & \multirow{2}{*}{GENERIC.MP\#8} \\
    & \\
    \cline{3-4}
    &  & \multirow{2}{*}{7.6}  &  \multirow{2}{*}{GENERIC.MP\#332} \\
    & \\
    \cline{2-6}
    & \multirow{5}{0.065\textwidth}{NetBSD} & \multirow{2}{*}{7.0} & \multirow{2}{0.15\textwidth}{GENERIC. 201509250726Z} & \multirow{5}{*}{IPv6/TCP} & \multirow{5}{0.4\textwidth}{IPv4 test cases with fragment sizes lower than 40 bytes do not receive any answer. We thus hypothesize the used patterns are not well suited to test NetBSD IPv4 policy. We do not observe such behavior with IPv6 or TCP. We do not observe any reassembly difference with npf firewall enabled/disabled.} \\
    &   \\
    \cline{3-4}
    &  & 8.0  & GENERIC\#0 &  & \\
    \cline{3-4}
    &  & \multirow{2}{*}{9.3} & \multirow{2}{*}{GENERIC\#0} &  & \\
    & \\
    \cline{2-6}
    & Solaris & 11.2 $\rightarrow$ 11.4 & SunOS 5.11 & IPv4/IPv6/TCP & Solaris' firewall is disabled by default, and we oberve IPv4 and IPv6 reassembly difference if enabled. \\
    \hline
    \multirow{4}{0.07\textwidth}{IoT/ embedded} & lwIP & 2.2.1 & - & IPv4/IPv6/TCP & - \\
    \cline{2-6}
    & picoTCP & 1.7.0 & - & IPv4/IPv6/TCP &  -\\
    \cline{2-6}
    & uIP & 1.0  & - & IPv6/TCP & IPv4 test cases are reassembled inconsistently across the tool runs. We thus could not extract the implementation policies. \\
    \cline{2-6}
    & smoltcp & 0.12.0 & - & IPv4/IPv6/TCP & We increased the default value of the REASSEMBLY\_BUFFER\_COUNT parameter from 1 to 2000. The default value leads to run inconsistencies for IPv4 tests. \\
    \hline
    Unikernel & mirage-tcpip & 9.0.0 & - & IPv4/IPv6/TCP & - \\
    \hline
    DPDK- compatible & Seastar & 22.11.0 & - & IPv4/TCP & Seastar does not implement IPv6. \\
    \hline
    \multirow{2}{*}{NIC} & Chelsio TOE & T520-CR & - & TCP & \multirow{2}{0.4\textwidth}{There is no IPv4 and IPv6 fragment offloading; therefore, fragments are forwarded as is to the OS kernel.} \\
    \cline{2-5}
     & Xilinx Onload & 9.0.1 & - & TCP & \\
    \hline
    \multirow{5}{*}{NIDS} & \multirow{2}{*}{Suricata} & \multirow{2}{*}{7.0.9}  & - & \multirow{2}{*}{IPv4/IPv6/TCP} & \multirow{4}{0.4\textwidth}{These NIDSes have configurable reassembly policies. \Cref{sec:uc_pair_triplet_reassembly_policies,sec:uc_3_custom_algo} report NIDS reassemblies with the \emph{default} policies. \Cref{sec:uc_reassembly_bugs,sec:uc_reassembly_policy_generalization_temptative_ids_inconsistencies} report NIDS reassemblies with the \emph{configured} policies.}  \\
    \\
    \cline{2-4}
    & \multirow{2}{*}{Snort} & \multirow{2}{*}{3.7.1.0}  & - & \multirow{2}{*}{IPv4/IPv6/TCP} & -\\
    \\
    \cline{2-6}
    & Zeek & 7.1.1 & - & IPv4/IPv6/TCP & This NIDS does not provide the configurable reassembly policies. \\
    \bottomrule
    \end{tabular}
\end{table*}

\subsection{Custom reassembly algorithm test}
\label{sec:appendices_custom_reassembly_algorithm_test}

\subsubsection{The pipeline}
\label{sec:appendices_custom_reassembly_algorithm_test_the_pipeline}

\begin{figure*}[!ht]
    \centering
    \includegraphics[width = 1\textwidth]{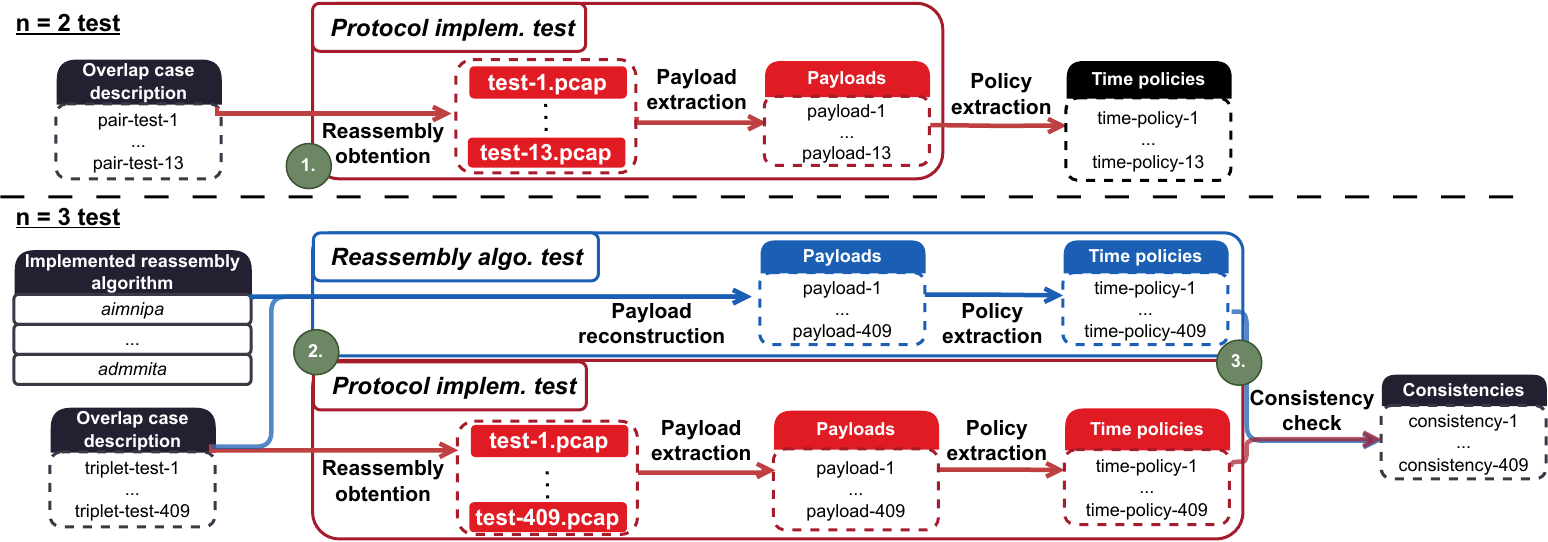}
    \caption{\label{images/reassembly_algo_pipeline}The pipeline for the reassembly algorithm test.
        This process is repeated for every (protocol, scenario, implementation) tuple.
        Black items are input or output elements.
    }
\end{figure*}

\Cref{images/reassembly_algo_pipeline} describes the overall pipeline used for the test of the implemented reassembly algorithms.
The pipeline is illustrated for a specific testing scenario, since pair and triplet time policies may change across scenarii.
In other words, the process is repeated for every scenario.

Let's take the $s^{sf,ef}$ and Debian 12 OS as an example for the IPv4 protocol.
First (cf \ding{172} on~\Cref{images/reassembly_algo_pipeline}), we run Debian 12 test in order to retrieve the $n=2$ time policies.
We get an association between individual Allen relations (or pair test cases, since $n=2$ test cases are composed of one unique Allen relation) and pair time policies, e.g., $Eq$ is associated to $Old$, $Fi$ to $Old$, $F$ to $Ignore$.
Second (cf \ding{173} on~\Cref{images/reassembly_algo_pipeline}), on one hand, we test \emph{individually} the implemented reassembly algorithms with $n = 3$ test cases, based on Debian 12's $n=2$ time policies; on the other hand, we test Debian 12 with $n = 3$ test cases.
Third (cf \ding{174} on~\Cref{images/reassembly_algo_pipeline}), this finally enable us to get the number of test case reassembly inconsistencies for every implemented algorithm by comparing their reassemblies with Debian 12.

\subsubsection{The observed reassembly mechanims with side effect}
\label{sec:appendices_reassembly_mechanisms_with_side_effect}

This part describes the reassembly mechanisms with side effect we observed within protocol implementation source codes. 
This manual investigation (temporarily) diverges from the opaque box approach described in~\Cref{sec:pyrolyse_design}.

\paragraph{Chunk merging for \{no, meet, any\} relations}

When there is no data hole between chunks, an implementation can merge them, for performance reason for example, as we've seen with Linux 6.1 in~\Cref{sec:chunk_merging}.
An implementation may want to merge only in some very specific situations.
That is why we implemented three merging strategies:
\begin{itemize}
    \item no chunk merging, i.e., \emph{no} strategy.
    \item chunk merging only for perfectly contiguous chunks, in other words $M$ or $Mi$ relations, i.e., \emph{meet} strategy.
    \item chunk merging for any Allen relation (except $B$ and $Bi$ relations), i.e., \emph{any} strategy.
\end{itemize}

\paragraph{Chunk characteristic \{immediate, delayed\} alteration during overlap resolution}

When a new chunk arrives and is compared with queued chunks before its (possible) insertion, the alteration of the newly arrived chunk's characteristics, i.e., beginning and ending data offsets, are modified either \emph{immediately} or \emph{delayedly}.
More precisely, if the chunk overlaps with several ones, either 1) its beginning and ending offsets are adjusted after every overlap comparison with already inserted chunks or 2) its characteristics stay unchanged for the remaining comparisons until its (possible) insertion.

\paragraph{Interpretation of the \emph{ignore} time policy as \{drop of all the triplet's chunks, drop of the pair's chunks,  drop of the oldest pair chunk, drop of the newest pair chunk\}}

An implementation can \emph{ignore} a test case without any hole by not answering back.
Because we treat the tested implementations as opaque boxes, we do not know exactly what caused the implementation not to respond to the test case.
For example, IPv6 specification~\cite{deering2017rfc} says that all the fragments that corresponds to the same original datagram should be silently discarded.
In the $n=3$ testing context, if the time policy of at least one of the relations describing a triplet test case is \emph{ignored} (based on $n=2$ testing), this could either mean that:
\begin{enumerate}
    \item [1.] all the test case chunks are dropped and therefore, the entire triplet test case is \emph{ignored}.
          In this case, we consider that no matter if $p_{01}$, $p_{02}$ or $p_{12}$'s time policy is \emph{ignore}, the information is stored and, therefore, all the following chunks are dropped.
          In other words, the flow is not reset (i.e., the implementation keeps the information that a pair has been ignored).
    \item [2.] only the corresponding pair's chunks are dropped.
          The flow is then reset.
          Therefore, if we consider a triplet where $p_{01}$ is ignored and a chunk $2$ with data starting at byte offset 0 and finishing the rightmost, the triplet test case can ultimately be reassembled with chunk $2$'s data.
\end{enumerate}
Finally, an implementation that \emph{ignores} the pair $p_{ij}$ in a triplet test case can also:
\begin{enumerate}
    \item [3.] drop the oldest, i.e., $i^{th}$, chunk.
    \item [4.] drop the newest, i.e., $j^{th}$, chunk.
\end{enumerate}

\subsubsection{List of reassembly algorithms}
\label{sec:implemented_algorithms}

This part list the IP and TCP reassembly algorithms we implemented and tested. 
In our case, a reassembly algorithm is basically a set of reassembly mechanism with side effet described in~\Cref{sec:appendices_reassembly_mechanisms_with_side_effect}.

\paragraph{IP protocols}

We implemented the 12 reassembly algorithms reported in~\Cref{table:ip_algo_acronyms}.

\renewcommand{\arraystretch}{1.3}
\begin{table}[t!]
    \setlength\tabcolsep{4pt}
    \centering
    \small

    \caption{\label{table:ip_algo_acronyms}Implemented IP reassembly algorithms.}
    \begin{tabular}{l l l l}
        \toprule

        \multirow{2}{*}{\textbf{Algorithm}} & \multicolumn{3}{c}{\textbf{Reassembly mechanism}}                                             \\
        \cline{2-4}
                                             & \textbf{\makecell[l]{characteristic\\\textcolor{red}{a}lteration}} & \textbf{\makecell[l]{relation\\\textcolor{red}{m}erging}} & \textbf{\makecell[l]{drop \emph{\textcolor{red}{i}gnore}\\interpretation}} \\
        \midrule
        \emph{aimnipa}                                                           & immediate & no   & all chunk pair    \\
        \emph{aimnipn}                                                           & immediate & no   & newest chunk pair \\
        \emph{aimnita}                                                           & immediate & no   & all chunk triplet \\
        \emph{admnita}                                                           & delayed   & no   & all chunk triplet \\
        \emph{aimaipa}                                                           & immediate & any  & all chunk pair    \\
        \emph{aimaipn}                                                           & immediate & any  & newest chunk pair \\
        \emph{aimaita}                                                           & immediate & any  & all chunk triplet \\
        \emph{admaita}                                                           & delayed   & any  & all chunk triplet \\
        \emph{aimmipa}                                                           & immediate & meet & all chunk pair    \\
        \emph{aimmipn}                                                           & immediate & meet & newest chunk pair \\
        \emph{aimmita}                                                           & immediate & meet & all chunk triplet \\
        \emph{admmita}                                                           & delayed   & meet & all chunk triplet \\

        \bottomrule
    \end{tabular}
\end{table}

We did not implement the \emph{oldest chunk pair drop} ignoring strategy because of its development effort.
And, we do not associate the \emph{delayed} chunk characteristic alteration with \emph{all chunk pair drop} and \emph{newest chunk pair drop} because such association does not always make sense.

\paragraph{TCP protocol}

We implemented the six reassembly algorithms reported in~\Cref{table:tcp_algo_acronyms}.

\renewcommand{\arraystretch}{1.3}
\begin{table}[t!]
    \centering
    \small

    \caption{\label{table:tcp_algo_acronyms}Implemented TCP reassembly algorithms.}
    \begin{tabular}{l lll}
        \toprule

        \multirow{2}{*}{\textbf{Algorithm}} & \multicolumn{2}{c}{\textbf{Reassembly mechanism}}                    \\
        \cline{2-3}
                                             & \textbf{\makecell[l]{characteristic\\\textcolor{red}{a}lteration}} & \textbf{\makecell[l]{relation\\\textcolor{red}{m}erging}}  \\
        \midrule
        \emph{aimn}                                                              & immediate & no   \\
        \emph{admn}                                                              & delayed   & no   \\
        \emph{aima}                                                              & immediate & any  \\
        \emph{adma}                                                              & delayed   & any  \\
        \emph{aimm}                                                              & immediate & meet \\
        \emph{admm}                                                              & delayed   & meet \\

        \bottomrule
    \end{tabular}
\end{table}

We did not implement the mechanism related to the $Ignore$ pair time policy interpretation, since we did not observe in practice such time policy for TCP.

\subsection{Detailed responsible disclosure process}
\label{sec:responsible_disclosure_process}

\Cref{tables/responsible_disclosure_summary} give details regarding the responsible process. 

\renewcommand{\arraystretch}{1.1}
\begin{table*}[t!]
    \caption{\label{tables/responsible_disclosure_summary}Responsible disclosure summary. Provided artifacts are composed of a technical report and the related PCAP or log materials.
    }
    \setlength\tabcolsep{4pt}
    \centering
    \begin{tabular}{L{0.15\textwidth} L{0.08\textwidth} L{0.11\textwidth} L{0.32\textwidth}  L{0.22\textwidth}}
    \toprule
    \textbf{Disclosure type} & \textbf{Date} & \textbf{Recipient} & \textbf{Description}  &   \textbf{Comments} \\
    \midrule
    \multirow{8}{0.15\textwidth}{Reassembly errors/ bugs} & 06/11/2023 & Suricata & IP reassembly logic error  & Partially fixed, credited with CVE~\cite{cve_ours} \\
    \cline{2-5} 
    & 06/11/2023  & Snort & IP reassembly logic error &  Not fixed \\
    \cline{2-5} 
    & 06/11/2023 & Suricata & TCP log display error &  Partially fixed \\
    \cline{2-5} 
    & 06/12/2024 & Suricata &  TCP reassembly logic error & Partially fixed \\
    \cline{2-5} 
    & 24/01/2025 & OpenBSD & IP reassembly logic error &  Fixed~\cite{openbsdpatchip} \\
    \cline{2-5} 
    & 28/01/2025 & OpenBSD & TCP connection termination error &  Fixed~\cite{openbsdpatchtcp} \\
    \cline{2-5} 
    & 27/03/2025 & lwIP & TCP connection termination error &  Not fixed \\
    \cline{2-5} 
    & 27/03/2025 & mirage-tcpip & TCP connection termination error &  Not fixed \\
    \hline
    \multirow{2}{0.15\textwidth}{($n=2$) pair policy update and extension} & 29/01/2024 & Suricata/Snort/ Zeek & Latest OS and NIDS pair reassembly policies obtained with the IP $s^{c}_{nf}$ and TCP $s^{sf}$ scenarii. It reported the NIDS reassembly inconsistencies with the latest OSes. & Suricata fixed its default, i.e., $bsd$, IP reassembly policy~\cite{cve_ours} \\
    \cline{2-5} 
    & 26/08/2024 & Suricata/Snort/ Zeek & OS pair reassembly policies obtained with the $S^{IP}$ and $S^{TCP}$ testing scenarii. It reported the reassembly diversity across testing scenarii.  &  No answer \\
     \hline
    \multirow{2}{0.15\textwidth}{($n=3$) triplet policy description} & 26/02/2025 & Suricata/Snort/ Zeek & Latest OS triplet reassembly policies obtained with the $S^{IP}$ and $S^{TCP}$ testing scenarii. It reported the reassembly inconsistency between ($n=2$) pairs and ($n=3$) triplets test cases. The custom reassembly algorithms were described and tested. & No answer\\
    \cline{2-5} 
    & 31/03/2025 & Suricata/Snort/ Zeek & Triplet reassembly policies of older OS versions, NIDS, and all the other network stacks. Policies are obtained with the $S^{IP}$ and $S^{TCP}$ testing scenarii. It reported the reassembly inconsistency between ($n=2$) pairs and ($n=3$) triplets test cases. The custom reassembly algorithms were described and tested. & No answer\\
    \bottomrule
\end{tabular}                                                                        
\end{table*}

\end{document}